\documentclass{JHEP3}
\usepackage{graphicx}
\usepackage{feynmp}
\usepackage{cite}

\newcommand{\tfrac}[2]{{\textstyle\frac{#1}{#2}}}

\newcommand{\eslsh}{\epsilon\kern-0.4em/}
\newcommand{\pslsh}{p\kern-0.5em/}
\newcommand{\spslsh}{p\kern-0.365em/}
\newcommand{\ppslsh}{p'\kern-0.75em/}
\newcommand{\kslsh}{k\kern-0.5em/}
\newcommand{\tppslsh}{\tilde{p}'\kern-0.75em/}
\newcommand{\tpslsh}{\tilde{p}\kern-0.5em/}
\newcommand{\poneslsh}{p\kern-0.5em/_1}
\newcommand{\ptwoslsh}{p\kern-0.5em/_2}
\newcommand{\pin}{p_{\mathrm{in}}}
\newcommand{\scdot}{{\cdot}}

\title{Collinearity, convergence and cancelling infrared divergences }
\author{Martin Lavelle and David McMullan\\School of Mathematics and Statistics\\
University of Plymouth\\Plymouth, PL4 8AA\\UK\\
E-mail: \email{mlavelle@plymouth.ac.uk}\quad
\email{dmcmullan@plymouth.ac.uk}}

 \abstract{The Lee-Nauenberg theorem is a fundamental quantum mechanical result which provides the
 standard theoretical response to the problem of collinear and infrared divergences. Its argument,
 that the divergences due to massless charged particles can be removed by summing over degenerate
 states, has been successfully applied to systems with final state degeneracies such as LEP processes.
 If there are massless particles in both the initial and final states, as will be the case at the LHC,
 the theorem  requires the incorporation of disconnected diagrams which produce connected
 interference effects at the level of the cross-section. However, this aspect of the theory has never
 been fully tested in the calculation of a cross-section. We show through explicit examples that in such
 cases the theorem introduces a divergent  series of diagrams and hence fails to cancel the infrared divergences.
 It is also demonstrated that the widespread
 practice of treating soft infrared divergences
 by the Bloch-Nordsieck method and handling collinear divergences by the Lee-Nauenberg method is not
 consistent in such cases.
}

\begin{document}

\section{Introduction}
Gauge theories have very good ultraviolet properties
\cite{'tHooft:1972fi} and, building upon this, have become the bed
rock  for the standard model of particle physics. In stark contrast
to this, though, they have very bad infrared behaviour  which
obstructs a deeper understanding of their large scale properties and
hence their predictive power.

The  infrared problem in gauge theories has a long history. It was
addressed, and to a limited sense  solved, in one of the earliest
papers on QED~\cite{Bloch:1937pw}. In that paper, Bloch and
Nordsieck recognised that in, for example, Coulombic scattering
there is always the possibility for the emission of a soft photon no
matter how far the electron is from the scattering event. The
fundamental reason for this is that the photon is massless, so there
is always enough energy to emit a photon with a suitably long
wavelength. Indeed, given the finite energy resolution of any
experiment, the processes of electron scattering and electron
scattering accompanied by the emission of an arbitrary number of
soft photons (so that overall their energy is less than the
experimental resolution) are indistinguishable. The inclusive
cross-section formed by summing over all soft final photons was
shown, by Bloch and Nordsieck, to be infrared finite but resolution
dependent.

The Bloch-Nordsieck mechanism for dealing with the infrared problem
in QED is physically appealing but not fully satisfying
theoretically. It did not explain why the virtual corrections to
tree-level Coulombic scattering suffered from infrared divergences
in the first place, but only that they could be cancelled against
the processes of emitting real but soft photons. There is also an
unattractive time asymmetry in the Bloch-Nordsieck argument: one is
forced to assume that the in-state is \emph{just} an electron while
the out-state has an electron \emph{and} the experimentally
unobservable soft photons. Surely, though,  the in-state could also
be contaminated with unobservable photons? However, including these
would introduce new soft divergences and it is not immediately clear
how they could be cancelled.

In the early sixties our understanding of the infrared properties of
QED was greatly advanced by the works of Lee and
Nauenberg~\cite{Lee:1964is} and Chung~\cite{Chung:1965}. These
authors  addressed different questions concerning the infrared and
hence had very little overlap in their conclusions. However, their
insights highlighted complementary aspects of the infrared that, we
feel, must ultimately be fused into a common strategy for dealing
with the large scale aspects of gauge theories.

On the face of it, the most conservative approach was that of Lee
and Nauenberg. They wanted to extend the Bloch-Nordsieck mechanism
to include the additional (collinear or mass) infrared structures
that arise when, for example, the mass of the electron is taken to
zero in Coulombic scattering. Building upon lessons learnt through
explicit calculations by Kinoshita~\cite{Kinoshita:1962ur}, they
were able to prove a rather general quantum mechanical result. When
applied to a field theory with massless fields, their formal
argument concludes that cross-sections are free of both soft and
collinear divergences if summed over both final \emph{and} initial
degenerate states. Degenerate means here degenerate in energy up to
the resolving power of any given experiment. It is important to
note, though, that they do not explicitly  investigate how soft
divergences cancel in this way and assume \lq\lq \emph{the infrared
divergence has already been eliminated by including the
contributions due to the emissions of soft photons}\rq\rq\ (see the
discussion following equation (20) in~\cite{Lee:1964is}).

Chung, on the other hand, was not concerned with additional infrared
structures. He wanted to go beyond the cross-section approach of
Bloch-Nordsieck and develop an infrared finite S-matrix description
of scattering in QED. The challenge he faced was that the virtual
processes introduced an infrared divergence at the S-matrix level,
while the emission of a soft photon only produces an infrared
contribution when integrated over in the cross-section. His response
to this  was to introduce a new type of process: the emission and
absorption of a very specific coherent state of photons. These
photons were built up of wave-packets smeared by functions that
contained the necessary infrared singularities to cancel the virtual
contribution. Note that \emph{both} emission and absorption are
essential here, as is the inclusion of the disconnected process
where the photon does not interact at all with the electrons (see
Figure~2 in~\cite{Chung:1965}). The fact that these enter at the
same order of perturbation theory as the virtual process clearly
illustrates the non-standard (coupling dependent) nature of the
coherent state wave packets.

Chung also talks about a resolution but now it refers to the range
in momenta that the infrared coherent state wave packets are
integrated over. Its role is to provide a divide between hard and
soft processes, and to this extent it is arbitrary. Through it,
though, one could argue that the coherent states ameliorated the
infrared while not altering the ultraviolet properties of the
theory.

The coherent state approach was considerably refined by
Kibble~\cite{Kibble:1968} and then later by Kulish and
Faddeev~\cite{Kulish:1970ut}. These works showed that the origin of
the soft infrared divergences could be traced back to the fact that
the asymptotic interaction in QED never vanishes due to the
masslessness of the photon. An asymptotic interaction picture could
be constructed, resulting in a theory of infra-particle scattering.
The infra-particle being identified as a distorted free particle,
where the field dependent distortion can be identified with Chung's
infrared coherent state.  A characteristic feature of this approach
is the abandonment of a  particle description. That is, the charges
are not identified with poles in the propagator, but rather with
branch cuts.

 The
non-trivial asymptotic dynamics associated with soft degrees of
freedom has been extended to include collinear
structures~\cite{Contopanagos:1991yb,Forde:2003jt}, at least to the
lowest non-trivial order in the coupling. It has also been formally
extended to include some aspects of the much more complicated
infrared properties found in non-abelian
theories~\cite{Nelson:1980yt}. Through this we see that the S-matrix
approach to the infrared problems, based on the recognition that
massless particles imply a non-trivial asymptotic dynamics, provides
a consistent formalism for addressing some key issues related to
both soft and collinear divergences. However, as it stands, this
approach is unattractive computationally  and physically since, in
particular, it relies on the use of old fashioned time-ordered
perturbation theory and the loss of a particle description of
charges~\cite{Kulish:1970ut}.

The Lee-Nauenberg approach to the infrared seems, at first sight,
less of a departure from the more conventional techniques  used in
particle physics. Certainly, the cancellation of collinear
divergences in Coulombic scattering~\cite{Lee:1964is} or pair
production (see, for example, Section 8.6 in~\cite{Brown}) is
 more familiar and simpler to calculate in the cross-sectional
approach (compare with ~\cite{Contopanagos:1991yb} and
\cite{Forde:2003jt}). This has made it a much more attractive way to
understand the infrared for the bulk of the particle physics
community. To quote Sterman (see page 443 of~\cite{Sterman:1994ce}):
\lq\lq \emph{For applications to high-energy scattering, its
importance has thus far been more conceptual than practical, but it
is a fundamental theorem of quantum mechanics and puts many specific
results in perspective.}\rq\rq\ Given this central role that it
plays in our understanding of the infrared, it is important that we
have a proper understanding of  how to use the Lee-Nauenberg theorem
in quantum field theory.

In this paper we want to see, through concrete examples, how the
Lee-Nauenberg theorem should  be  used in practice in field theory.
In their original paper, Lee and Nauenberg used, at least from a
modern point of view, some  non-standard techniques that obscure a
full understanding of their method and mask some important
consequences of this approach to the infrared. Although several
authors~\cite{Muta:1981pe,Axelrod:1985yi,Akhoury:1997pb} have
revisited some aspects of the arguments used by Lee and Nauenberg,
there has not been, to the best of our knowledge, a systematic
reappraisal of how their method should be applied to gauge theories
when there are both initial and final state degeneracies. Given the
relevance of precisely this type of process to the forthcoming LHC
era in particle physics, such a reassessment of the role of the
Lee-Nauenberg theorem is, we feel, particularly timely.

In the original paper~\cite{Lee:1964is}, two examples were presented
to illustrate their mechanism for collinear cancellations. Their
work was restricted to the first few degenerate diagrams that arise
in these processes. In this paper we plan to present these two
examples again, but now in full. That is, we will not arbitrarily
truncate the Lee-Nauenberg method, but allow for the full degeneracy
required in their theorem. The conclusions of this will be quite
striking: we will see that there is an incompatibility in these
processes between the Bloch-Nordsieck approach to the soft infrared
and the Lee-Nauenberg treatment of the collinear regime. We will
then explain how  a mass-resummation is needed to correctly
implement the Lee-Nauenberg proposal, a point not seen in their
original presentation. Finally, we will expose a problem with the
convergence of the expected infrared cancellation claimed by the
theorem when there are final and initial state degeneracies.

In Section~2 we  discuss Coulombic scattering as the mass of the
electron becomes small. This example was the main application
considered in~\cite{Lee:1964is}. We include it here to both fix
notation and to point out an inconsistency in the way soft and
collinear divergences were dealt with by Lee and Nauenberg. In
Section~3 we will treat the example of Coulombic scattering
accompanied by an observed out-going jet of collinear photons. Here
we will consider, as Lee-Nauenberg did,  one out-going photon
collinear with the in-coming electron. We will see why a mass
resummation is needed in order for Lee-Nauenberg's claims to be
implemented and then  start to explore the role of disconnected
processes in the cancellation of collinear divergences. In Section~4
we will allow for an arbitrary number of photons in the final jet.
The combinatorics that arises will be clarified and we will see how
collinear divergences are meant to cancel for such full jets. The
techniques developed will then, in Section~5, be applied to soft
divergences and we will identify the mechanism for soft cancellation
in the Lee-Nauenberg approach. Finally, in Section~6, we will focus
on the thorny issue of the convergence of the various series in
terms of the number of  photons. We will see that the series are, in
fact, divergent. The cancellation of collinear or soft divergences
only arises, as will be discussed, for a very specific ordering of
the respective series. We will conclude with a list of open problems
related to the implementation of the Lee-Nauenberg theorem in
quantum field theory.

\section{Degeneracies in Coulomb scattering}
\begin{fmffile}{coulomb}
Coulombic scattering of an electron is a basic process in field
theory and is the main example discussed by Lee and Nauenberg. We
will follow their presentation  by first considering the scattering
of massive electrons and then investigate the high energy limit
where the mass, $m$,  can effectively be taken to zero. \FIGURE{
\mbox{\begin{fmfgraph*}(30,30)
      \fmfleft{source}
      \fmf{photon,l.d=2cm}{source,ver}
      \fmf{plain}{v1,ver}
      \fmf{fermion,label=$p$,label.side=right}{pone,v1}
      \fmf{plain}{ver,v2}
      \fmf{fermion,label=$p'$,label.side=right}{v2,ptwo}
      \fmfdot{ver}
      \fmfright{pone,ptwo}
    \end{fmfgraph*}}
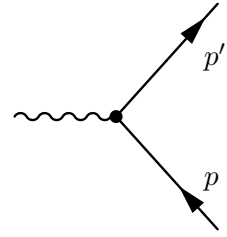
\caption{Coulombic scattering at tree-level.}} To this end we
consider the process where an in-coming electron of momentum $p$ is
scattered off a  nucleus and becomes an out-going electron with
momentum $p'$. We will always work in the lab frame, where the
nucleus is taken to be static, unless otherwise stated.

 The tree-level cross-section, as described in
Figure~1, is given by
\begin{equation}\label{coul1}
    \frac{d\sigma_0}{d\Omega}=\frac{\alpha^2}{|\underline{q}|^4}|\bar{u}'\gamma_0u|^2,\,
\end{equation}
where we write $\bar{u}'=\bar{u}(p')$ and $u=u(p)$, and there is an
implied sum over final spins and average over initial spins. Note
that, in order to keep track of in-coming and out-going particles,
we adopt the convention in this paper that all out-going particle
momenta are primed.

As is well known, after the usual ultraviolet renormalisation, the
one-loop correction to this process  contains two terms
characterised by the structure functions $F_1$ and $F_2$. The
function $F_2$ is responsible for the anomalous magnetic moment of
the electron and is infrared safe, so we will neglect it. The other
structure function is much more important to us and results in the
replacement of $\gamma_0$ in the  tree-level S-matrix by
$\gamma_0(1+F_1)$. Thus at the cross-section level we have
\begin{equation}\label{coul2}
    \frac{d\sigma_0}{d\Omega}\to\frac{d\sigma_0}{d\Omega}(1+2F_1)\,,
\end{equation}
where the factor of 2 arises from the two cross terms that
contribute at this order in the coupling. The key point to note is
that $F_1$ is singular in the infrared. Using dimensional
regularisation in $D=4+2\varepsilon_{_\mathrm{IR}}$ dimensions to
regulate the soft infrared divergences, we have
\begin{equation}\label{coul3}
2F_1=\frac{e^2}{4\pi^2}\left[-\frac1{\hat{\varepsilon}_{_\mathrm{IR}}}\left(\!\ln\left(\!
\frac{\!Q^2}{m^2}\right)-1\right)
+\frac12\ln^2\!\left(\!\frac{\!Q^2}{m^2}\right)+\frac12\ln\!\left(\!\frac{\!Q^2}{m^2}\right)
\left(1-2\ln\!\left(\!\frac{\!Q^2}{\mu^2}\right)\right)\right]\,,
\end{equation}
where $Q^2=-(p'-p)^2$,
$1/\hat{\varepsilon}=1/\varepsilon+\gamma-\ln4\pi$ and $\mu$ is the
mass scale that enters dimensional regularisation. In this
expression for $F_1$ we have neglected any terms that are finite as
either  $\varepsilon_{_\mathrm{IR}}\to0$ or $m\to0$.

The soft infrared divergence can be eliminated using the
Bloch-Nordsieck argument by including the degenerate process of soft
emission by photons where the photon energy is less than the energy
resolution $\Delta$ of the detector. \FIGURE{
    \parbox{30mm}{\begin{fmfgraph*}(30,35)
    \fmfleft{source}
     \fmftop{tip1}
      \fmf{photon,l.d=2cm}{source,ver}
      \fmf{plain}{v1,ver}
      \fmf{fermion,label=$p$,label.side=right}{pone,v1}
      \fmf{plain}{ver,v2}
      \fmf{fermion,label=$p'$,label.side=right}{v2,ptwo}
      \fmfright{pone,ptwo}
      \fmffreeze
      \fmfshift{(.7cm,0)}{tip1}
    \fmf{photon,label=$k'$,label.side=left}{v2,tip1}
    \fmfdot{ver,v2}
    \end{fmfgraph*}}
    +
\parbox{30mm}{\begin{fmfgraph*}(30,35)
      \fmfleft{source}
     \fmftop{tip1}
      \fmf{photon,l.d=2cm}{source,ver}
      \fmf{plain}{v1,ver}
      \fmf{fermion,label=$p$,label.side=right}{pone,v1}
      \fmf{plain}{ver,v2}
      \fmf{fermion,label=$p'$,label.side=right}{v2,ptwo}
      \fmfright{pone,ptwo}
      \fmffreeze
      \fmfshift{(.7cm,-1.7cm)}{tip1}
    \fmf{photon,label=$k'$,label.side=right}{v1,tip1}
    \fmfdot{ver,v1}
    \end{fmfgraph*}}
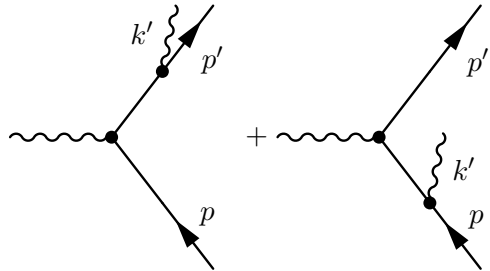
\caption{Soft emission from the out-going and in-coming electron.}}

\noindent Let us recall how this soft infrared cancellation works in
practice. The two diagrams shown in Figure~2 contribute to this
process at this order in perturbation theory. Note that for this
soft emission we can take both $p'$ and $k'$ to be on-shell. The
overall out-going momentum is then $p'+k'$ which, as $k'$ is soft,
we take to be degenerate with the tree-level process.

\medskip

The S-matrix for this is
\begin{equation}\label{coul3.1}
    -ie^2 \bar{u}'\left(\frac{2
    p'\scdot\epsilon'+\eslsh'\kslsh'}{2 p'\scdot k'}\gamma_0-\gamma_0\frac{2
    p\scdot\epsilon'-\kslsh'\eslsh'}{2 p\scdot k'}\right)u\,,
\end{equation}
where $\epsilon'=\epsilon(k',\lambda')$, with $\lambda'$ being the
polarisation label for the photon. Now, to simplify the extraction
of the soft infrared divergence,  we are free to drop the $\kslsh'$
terms in the numerator and hence arrive at the familiar eikonal
expression for the S-matrix elements:
\begin{equation}\label{coul3.2}
    -ie^2 \left(\frac{
    p'\scdot\epsilon'}{p'\scdot k'}-\frac{
    p\scdot\epsilon'}{ p\scdot k'}\right)\bar{u}'\gamma_0u\,.
\end{equation}
Here we recognise the conserved current
\begin{equation}\label{coul3.3}
  J^\mu(k')=\frac{p'^\mu}{p'\scdot k'}-\frac{p^\mu}{p\scdot k'}\,,
\end{equation}
associated with the classical change in momentum of the charge.
 In the cross-section we
need to sum over the photon's polarisations and, due to the gauge
invariance of the current~(\ref{coul3.3}), this results in the
replacement of $\epsilon'^\mu\epsilon'^\nu$ by $-g^{\mu\nu}$. Hence,
by integrating over the physically undetectable soft photons, we
arrive at the  inclusive cross-section
\begin{equation}\label{coul4}
   \frac{d\sigma_0}{d\Omega}(1+2F_1+R_s)\,,
\end{equation}
where, up to soft and collinear finite terms, the contribution of
real soft emitted photons is
\begin{eqnarray}\label{coul5}
R_s&=&e^2\int\limits_{\mathrm{soft}}\frac{d^{D-1}k'}{(2\pi)^22\omega'}\left(\frac{2p'\scdot
p}{p'\scdot k'p\scdot k'}- \frac{p'^2}{(p'\scdot
k')^2}-\frac{p^2}{(p\scdot k')^2}\right)\\
&=&\frac{e^2}{4\pi^2}\left[\frac1{\hat{\varepsilon}_{_\mathrm{IR}}}\left(\ln\!\left(\!
\frac{Q^2}{m^2}\right)-1\right)
-\frac12\ln^2\!\left(\!\frac{Q^2}{m^2}\right)+\ln\!\left(\!\frac{Q^2}{m^2}\right)
\left(1+2\ln2+\ln\!\left(\!\frac{\tilde{\Delta}^2}{\mu^2}\right)\right)\right].\nonumber
\end{eqnarray}
Comparing this with (\ref{coul3}), we see that the inclusion of this
degenerate process  cancels the soft divergences and also the double
logs of the collinear, mass singularities. We are thus left with the
collinearly divergent terms
\begin{equation}\label{coul6}
    \frac{e^2}{\pi^2}\ln\left(\frac{Q}m\right)\left[\frac34+\ln2-\ln\left(\frac{Q}{\tilde{\Delta}}\right)\right]\,.
\end{equation}
Note that in arriving at this expression we have worked in the Breit
frame. Hence it is the resolution $\tilde{\Delta}$ in that frame
that arises in (\ref{coul6}). To translate back into the lab frame
we note that in the Breit frame
\begin{equation}\label{coul6.01}
    Q=2\tilde{E}\sqrt{1-\frac{m^2}{\tilde{E}}}\,,
\end{equation}
where $\tilde{E}$ is the electron's energy in the Breit  frame.
Hence, since
\begin{equation}\label{coul6.1}
    \frac{\tilde{E}}{\tilde{\Delta}}=
    \frac{E\sin(\tfrac12\phi)}{\Delta\sin(\tfrac12\phi)}=\frac{E}{\Delta}\,,
\end{equation}
where $\phi$ is the scattering angle for the electrons in the lab
frame, we see that the residual collinear divergence describe
by~(\ref{coul6}) contributes to the cross-section as
\begin{equation}\label{coul6.2}
    \frac{e^2}{\pi^2}\ln\left(\frac{E}m\right)\left[\frac34-
    \ln\left(\frac{E}{\Delta}\right)\right]\frac{d\sigma_0}{d\Omega}\,,
\end{equation}
where we have dropped collinear finite terms. Here we see the main
content of the Bloch-Nordsieck method: by summing over out-going
processes that are degenerate in energy with the scattering of
massive electrons, we arrive at an infrared finite but resolution
dependent cross-section.

The masslessness of the photon also means that one could have
non-detectable in-coming soft photons. They are not included in the
Bloch-Nordsieck approach but one could easily construct an
alternative procedure where one summed over initial photons that
were absorbed by the electron and neglected the out-going photons.
This would also be infrared finite if the ensemble of initial
photons had the same distribution as the out-going ones. However, it
is not so obvious how including both initial \emph{and} final soft
photons would lead to an infrared finite result. Although, in
principle, covered by the Lee-Nauenberg theorem, this point is not
addressed by~\cite{Lee:1964is}, so we shall postpone a detailed
discussion of this until later and press on with the collinear
divergence in equation~(\ref{coul6.2}) that arise as $m\to0$.


In the high energy limit, the mass of the electron becomes
negligible and we have a new class of degenerate processes where the
emitted photon can be nearly parallel to the out-going electron. In
this collinear configuration, given a finite angular resolution
$\delta$ in our detector, we can only measure the total out-going
energy $E$. We do not know how it is distributed between the
electron and any degenerate photons. Given that we have already
integrated over soft photons, this means that we should include
collinear photons with energy from the resolution $\Delta$ to the
total energy $E$. We call these semi-hard collinear photons.

\FIGURE{
    \parbox{30mm}{\begin{fmfgraph*}(30,35)
    \fmfleft{source}
     \fmftop{tip1}
      \fmf{photon,l.d=2cm}{source,ver}
      \fmf{plain}{v1,ver}
      \fmf{fermion,label=$p$,label.side=right}{pone,v1}
      \fmf{plain}{ver,v2}
      \fmf{fermion,label=$p'_1$,label.side=right}{v2,ptwo}
      \fmfright{pone,ptwo}
      \fmffreeze
      \fmfshift{(.7cm,0)}{tip1}
    \fmf{photon,label=$k'$,label.side=left}{v2,tip1}
    \fmfdot{ver,v2}
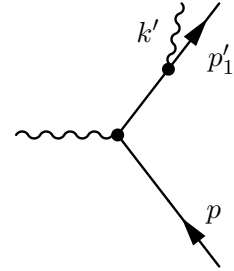
    \end{fmfgraph*}}\caption{Collinear emission from the out-going electron.
    }}

The emission of a photon collinear with the out-going electron can
take place from either electron line. However, it only forces an
internal line to go on-shell, and  potentially causes an infrared
divergence, if it is emitted from the out-going electron. So we only
need to consider the process where the photon of momentum $k'$ is
emitted from the out-going electron that ends up with momentum
$p'_1$, where $p'_1+k'=p'$. For semi-hard collinear photons, this
means that $k'$ and $p'_1$ are on-shell, but $p'$ is not. We denote
the energy of the out-going electron by $E_1$ where $E_1=E-\omega'$.
The S-matrix for this can be read off directly from Figure~3 to give
\begin{equation}\label{coul7}
    -i\frac{e^2}{2 p'_1\scdot k'}\bar{u_1}'(2
    p'_1\scdot\epsilon'+\eslsh'\kslsh')\gamma_0u\,,
\end{equation}
where now $\bar{u_1}'=\bar{u}(p'_1)$.

This contributes to the cross-section the term
\begin{equation}\label{coul8}
\frac{e^4}{(p'_1\scdot
k')^2}\left[4(p'_1\scdot\epsilon')^2(p'_1\scdot\tilde{p})-4(p'_1\scdot\epsilon')(\tilde{p}\scdot\epsilon')(p'_1\scdot
k')-2(\epsilon'\scdot\epsilon')(\tilde{p}\scdot k')(p'_1\scdot
k')\right]\,,
\end{equation}
where $\tilde{p}$ is defined by $\tpslsh=\gamma_0\pslsh\gamma_0$ so
that $4\tilde{p}\scdot p'=|\bar{u}'\gamma_0u|^2$. Prior to
integrating over the semi-hard jet of photons, we need to sum over
the photon polarisations. In contrast to the situation that arises
in the soft case where we could simply replace
$\epsilon'_\mu\epsilon'_\nu$ by $-g_{\mu\nu}$, we now need to use
the more general identity that
\begin{equation}\label{coul9}
   \sum_{\lambda'}\epsilon'_\mu\epsilon'_\nu=-g_{\mu\nu}-\frac{k'_\mu
   k'_\nu}{\omega'^2}+\frac{k'_\mu \eta_\nu+k'_\nu
   \eta_\mu}{\omega'}\,,
\end{equation}
with $\eta$ being the unit time-like vector so that, e.g.,
$\eta\scdot p'=\eta\scdot p=E$. Note that in~\cite{Lee:1964is} the
calculation  used an explicit helicity basis for the spinors. Here
we will see that the use of~(\ref{coul9}) simplifies the
calculations.

After summing over polarisations in (\ref{coul8}), we get
\begin{equation}\label{coul10}
    \frac{e^4}{p'_1\scdot k'}\left[\left(1+\frac{2E_1}{\omega'}\right)4(p'_1\scdot\tilde{p})+
    \left(1+\frac{E_1}{\omega'}\right)4(k'\scdot\tilde{p})\right]\,,
\end{equation}
where we have dropped all collinear finite terms. As discussed in
Appendix~B, for collinear photons and electrons we can, up to
collinear finite terms, write
\begin{equation}\label{coul11}
k'\scdot\tilde{p}=p'\scdot\tilde{p}\,\frac{\omega'}{E}\qquad\mathrm{and}\qquad
p'_1\scdot\tilde{p}=p'\scdot\tilde{p}\,\frac{E_1}{E}\,.
\end{equation}
Hence (\ref{coul10}) becomes
\begin{equation}\label{coul12}
 e^4|\bar{u}'\gamma_0u|^2\frac{E_1^2+E^2}{(p'_1\scdot
 k')E\omega'}\,.
\end{equation}
We can now use this to build up the inclusive cross-section for the
emission of collinear photons. In doing this it is important to note
that, since the emitted photon is now not soft, the in-coming and
out-going electrons have different energies. This means that in the
cross-section we must include the energy weighting $E_1/E$ familiar
from the Bethe-Heitler description of bremsstrahlung (see, for
example, Section 5-2-4 in \cite{IZ}, or the discussion of energy
weighting on page 499 of \cite{Brown}). The resulting cross-section
is given by
\begin{equation}\label{coul13}
     e^4|\bar{u}'\gamma_0u|^2\int\limits_{\shortstack{\scriptsize
     semi-hard\\\scriptsize cone}}^{ }\!\!\!\!\!\!\frac{d^3k'}{(2\pi)^32\omega'}
     \frac{E_1^2+E^2}{(p'_1\scdot
 k')E\omega'}\frac{E_1}E\,,
\end{equation}
where we write
\begin{equation}\label{coul13.1}
    p'_1\scdot k'=\tfrac12\omega'
    E_1\left(\theta_1^2+\frac{m^2}{E_1^2}\right)\,,
\end{equation}
and $\theta_1$ is the (small) angle between the out-going electron
and photon in the lab frame.

 Performing the angular integration over the cone with opening
angle $\delta$, and dropping collinear finite terms,  we get
\begin{equation}\label{coul14}
\frac{e^4}{4\pi^2}|\bar{u}'\gamma_0u|^2\ln\left(\frac{
E\delta}{m}\right)\frac1{E^2}\int_{\Delta}^E\frac{E_1^2+E^2}{\omega'}d\omega'\,.
\end{equation}
This final integral can be readily evaluated to yield the
cross-section for semi-hard collinear emission:
\begin{equation}\label{coul15}
    -\frac12\frac{e^2}{\pi^2}\ln\left(\frac{
E\delta}{m}\right)\left[\frac34-\ln\left(\frac{E}{\Delta}\right)-
\frac{\Delta}{E}+\frac14\frac{\Delta^2}{E^2}\right]\frac{d\sigma_0}{d\Omega}\,.
\end{equation}
Comparing this result with (\ref{coul6.2}) we see that just
including the out-going semi-hard collinear photons does not
completely remove the residual collinear divergences. In particular,
the pre-factor of a half found in (\ref{coul15}) obstructs  the
cancellation of the divergent terms found in the Bloch-Nordsieck
result~(\ref{coul6.2}).  However, following Lee-Nauenberg, if we now
include the in-coming degenerate process whereby a semi-hard photon
is absorbed by the in-coming electron, then we will get another
contribution equal to~(\ref{coul15}) and we thus see the
cancellation of the mass logarithms in~(\ref{coul6.2}). This was the
conclusion reached by Lee and Nauenberg following their
equation~(21).

However, there are still collinear divergent terms in (\ref{coul15})
that are linear and quadratic in $\Delta/E$ and these must be
cancelled  in order to get a collinear finite cross-section.  To
trace what is going on here, we note that these terms come from the
semi-hard energy integral in~(\ref{coul14}). Indeed the energy
integral splits into two terms:
\begin{equation}\label{coul16}
    \int_{\Delta}^E\frac{E_1^2+E^2}{\omega'}d\omega'=2E^2\int_{\Delta}^E\frac{d\omega'}{\omega'}+
    \int_{\Delta}^E(\omega'-2E)d\omega'\,.
\end{equation}
In the first term it is essential that the lower limit of $\Delta$
is kept otherwise we would reintroduce the soft divergences.
However, the second term is finite as $\Delta\to0$ and hence we see
that what is missing is a soft-collinear contribution from the
emitted photon that is finite in the soft regime. Thus we conclude
that the separation between soft and semi-hard photons is not a
precise division between soft and collinear divergent structures.

It is, in fact, not too difficult to trace where such a term was
dropped in the  Bloch-Nordsieck mechanism. In the discussion
following equation~(\ref{coul3.1}) we made the normal soft
simplification of dropping factors of $k'$ in the numerator. This,
however, has thrown away a relevant collinear term. Indeed,
precisely  the corresponding term in~(\ref{coul7}) generates the
divergent terms  in~(\ref{coul15}). Reinstating this momentum
in~(\ref{coul3.2}) and integrating the energy from 0 to $\Delta$, we
see that~(\ref{coul6.2}) should be replaced with
\begin{equation}\label{coul17}
    \frac{e^2}{\pi^2}\ln\left(\frac{E}m\right)\left[\frac34-
    \ln\left(\frac{E}{\Delta}\right)+\frac{\Delta}{2E}-\frac{\Delta^2}{8E^2}\right]\frac{d\sigma_0}{d\Omega}\,.
\end{equation}
This represents the  Bloch-Nordsieck analysis with all collinear
terms retained. Now we see that the consistent combination of the
Bloch-Nordsieck treatment of soft photon emission with the emission
of semi-hard collinear photons results in the cross-section
\begin{equation}\label{coul18}
    \frac12\frac{e^2}{\pi^2}\ln\left(\frac{E}m\right)\left[\frac34-
    \ln\left(\frac{E}{\Delta}\right)\right]\frac{d\sigma_0}{d\Omega}\,.
\end{equation}
As expected, this is still collinearly divergent and we need to
include the contribution from initial degenerate states. Now,
though, we face a problem not addressed in \cite{Lee:1964is}. We
have seen that soft initial states are ignored in  the
Bloch-Nordsieck analysis, and thus there is no equivalent consistent
procedure for including initial soft photon contributions that are
not infrared divergent. Hence we are forced to simply add the
in-coming version of~(\ref{coul15}) resulting in the cancellation
of~(\ref{coul18}) but the retention of the linear and quadratic
terms in~(\ref{coul15}). This means that, in such an approach to
collinear divergences, we must include in the integral over initial
states soft photon contributions which produce collinear divergences
but ignore those which generate soft divergences. This is extremely
unnatural.

Thus we see that, contrary to the procedure in~\cite{Lee:1964is}, we
cannot in general\footnote{It should be noted, though, that this
problem does not arise if there are only massless particles in the
final state since then the soft and collinear structures are dealt
with in a consistent manner.}  separately treat the soft and
collinear divergences using a mixture of Bloch-Nordsieck and
Lee-Nauenberg arguments. This does not mean that the general
Lee-Nauenberg theorem is wrong. What is does mean is that we need to
understand how to consistently deal with both soft and collinear
initial and final state degeneracies. As we have seen, a naive
inclusion of absorption from initial soft photons will double the
infrared divergences that arise from real processes and hence lose
the Bloch-Nordsieck soft cancellation. This suggests that the
mechanism for infrared cancellations is more subtle than expected.

In order to understand how we must refine our use of the
Lee-Nauenberg method for dealing with the infrared, we now look at
the second example discussed briefly in Appendix~D
of~\cite{Lee:1964is}.  We will now see how this application exposes
in a much deeper way the key  ingredient of their method when there
are both final \emph{and} initial degeneracies.

\end{fmffile}
\section{Coulomb scattering accompanied by a collinear jet}

\begin{fmffile}{collinear}

In addition to the collinear structures that  arise  in Coulombic
scattering as the mass of the electron becomes negligible, Lee and
Nauenberg also considered the process described in Figure~4, where
we have Coulombic scattering  accompanied by the emission of a
photon collinear with the incoming electron.
\FIGURE{\parbox{10cm}{\begin{center}\includegraphics{jets.1}
\end{center}
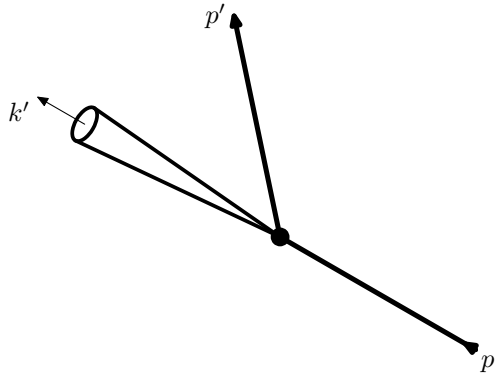
\caption{A jet of collinear photons emitted from the incoming
electron.}}}The resulting jet of photons of momentum $k'$ in the
final state would be clearly distinguished from the outgoing
electron as long as a scattering from the target has taken place,
such  that there is a large (measurable) angle between $p'$ and
$k'$, and the energy of the photon is greater than the experimental
resolution.

The technical attraction  of this process is that the collinear
structures that we wish to study already arise at tree-level. This
clearly simplifies the identification of the collinear and
degenerate aspects of the process. However, even here, the
identification of the collinear divergent terms is not
straightforward.  Here we will present for the first time a
comprehensive and more modern approach to this process.

\FIGURE{\parbox{3cm}{
\begin{fmfgraph*}(30,35)
      \fmfleft{source}
     \fmftop{tip1}
      \fmf{photon,l.d=2cm}{source,ver}
      \fmf{plain}{v1,ver}
      \fmf{fermion,label=$p$,label.side=right}{pone,v1}
      \fmf{plain}{ver,v2}
      \fmf{fermion,label=$p'$,label.side=right}{v2,ptwo}
      \fmfright{pone,ptwo}
      \fmffreeze
     \fmfshift{(.2cm,-1.7cm)}{tip1}
    \fmf{photon,label=$k'$,label.side=right}{v1,tip1}
    \fmfdot{v1,ver}
    \end{fmfgraph*}}
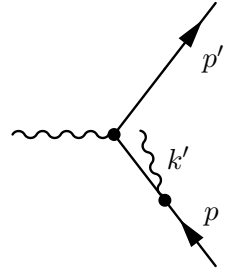
\caption{A single collinear photon emitted from the incoming
electron.}\label{fig5}}

The lowest order diagram that contributes to this is given in
Figure~\ref{fig5}. Its contribution to the S-matrix is:
\begin{equation}\label{s1}
    -\frac{ie^2}{(p-k')^2-m^2}\bar{u}'\gamma_0(\pslsh-\kslsh'+m)\eslsh'
   u=\frac{ie^2}{2p\scdot k'}\bar{u}'\gamma_0(\pslsh-\kslsh'+m)\eslsh'
   u\,,
\end{equation}
where $p$,  the incoming electron's momenta, is on-shell.

 As in all such processes, infrared
divergences arise due to the possibility that an internal line can
go on-shell for some choices of the external momenta. Here we see
that this occurs  when $p\scdot k'$ vanishes. Since $k'$ is not soft
but  the electron's mass is negligible, this happens if  the angular
separation between  $k'$ and $p$ is vanishingly small.

 We can now  identify the associated collinearly divergent
contribution to the cross-section associated with this scattering.
The resulting cross-section is
\begin{equation}\label{s2}
    \frac{2e^4}{(p\scdot k')^2}\left\{2(p\scdot\epsilon')^2(\tilde{p}'\scdot p-\tilde{p}'\scdot k')
    +2(p\scdot\epsilon')(\tilde{p}'\scdot\epsilon')(p\scdot
    k')-2(\epsilon')^2(\tilde{p}'\scdot k')(p\scdot k')
    \right\}
\end{equation}
and we note that here the energy weighting is trivial as the
in-coming and out-going electrons both have energy $E$.

After summing over photon polarisations, we arrive at the
cross-section
\begin{equation}\label{s4}
    \frac{e^4}{p\scdot k'}\left\{\left(\frac{2E}{\omega'}-1\right)4(\tilde{p}'\scdot p)+
    \left(1-\frac{E}{\omega'}\right)4(\tilde{p}'\scdot k')\right\}\,.
\end{equation}
If we now make the collinearity approximations~(\ref{coul11}), we
rapidly arrive at the expression
\begin{equation}\label{s7}
P_{0,1}(k',p):=e^4|\bar{u}'\gamma_0
u|^2\frac{2E_1E_0+\omega'^2}{(p\scdot k')E\omega'}\,.
\end{equation}
Here we are modifying the notation introduced
in~\cite{Akhoury:1997pb} and identifying  the Lee-Nauenberg
probability $P_{0,1}(k',p)$ as the contribution to the cross-section
with no in-coming photons but one out-going photon.

 \FIGURE{
\begin{fmfgraph*}(30,35)
    \fmfleft{source}
     \fmftop{tip1}
      \fmf{photon,l.d=2cm}{source,ver}
      \fmf{plain}{v1,ver}
      \fmf{fermion,label=$p$,label.side=right}{pone,v1}
      \fmf{plain}{ver,v2}
      \fmf{fermion,label=$p'$,label.side=right}{v2,ptwo}
      \fmfright{pone,ptwo}
      \fmffreeze
      \fmfshift{(-.6cm,0cm)}{tip1}
    \fmf{photon,label=$k'$,label.side=left}{v2,tip1}
    \fmfdot{v2,ver}
    \end{fmfgraph*}
  \caption{Emission of a photon from the out-going electron.}}

Having identified a collinearly divergent process, we now need to
find degenerate processes which also diverge and then sum them up.
The most obvious process to consider would be the emission of a
photon from the out-going electron that is collinear with the
in-coming one, as shown  in Figure~6.

However, as long as $k'$ is not soft and is parallel to $p$ rather
than $p'$, this does not force any internal line in this diagram to
go on-shell and hence does not lead to a divergent cross-section.
What is needed are degenerate processes where the emitted collinear
photon comes from the in-coming electron.

\FIGURE{\parbox{10cm}{\begin{center}\includegraphics{jets.2}
\end{center}
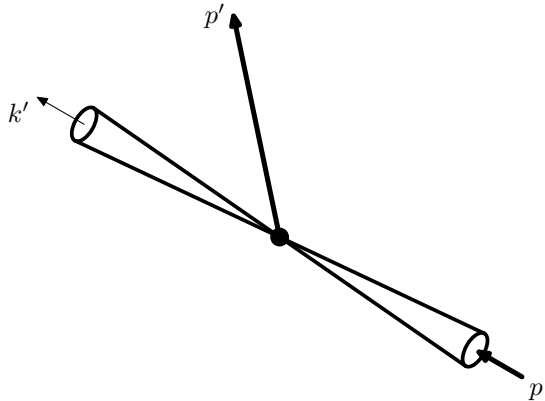
\caption{A jet of collinear photons emitted from the incoming
charged jet.}}}

 Given that we are working at order $e^4$ in the
coupling, it is not at first obvious that there are any such
degenerate processes contributing to the cross-section. However, the
 Lee-Nauenberg theorem says  we need also to
consider the initial state degeneracies as described in Figure~7. So
the  in-coming electron itself should be viewed as a degenerate
state formed from a mixture of the electron and collinear photons.

For example, we could consider the process whereby an initial photon
is first absorbed and then emitted from the in-coming electron or
vice-versa as shown in Figure~8.
 \FIGURE{\parbox{8cm}{\begin{center}
\parbox{30mm}{\begin{fmfgraph*}(30,35)
    \fmfleft{source}
     \fmfbottom{tip2,tip1}
      \fmf{photon,l.d=2cm}{source,ver}
      \fmf{plain}{v1,ver}
      \fmf{plain}{v12,v1}
      \fmf{fermion,label=$p_1$,label.side=right}{pone,v12}
      \fmf{plain}{ver,v2}
      \fmf{plain}{v21,v2}
      \fmf{fermion,label=$p'$,label.side=right}{v21,ptwo}
      \fmfright{pone,ptwo}
      \fmffreeze
      \fmfshift{(-1.2cm,-.3cm)}{tip1}
    \fmf{photon,label=$k$,label.side=right}{v12,tip1}
     \fmfshift{(2cm,1.5cm)}{tip2}
    \fmf{photon,label=$k'$,label.side=right}{v1,tip2}
    \fmfdot{v1,ver,v12}
    \end{fmfgraph*}}
    \quad + \quad
\parbox{30mm}{\begin{fmfgraph*}(30,35)
      \fmfleft{source}
     \fmfbottom{tip1,tip2}
      \fmf{photon,l.d=2cm}{source,ver}
      \fmf{plain}{v1,ver}
      \fmf{plain}{v12,v1}
      \fmf{fermion,label=$p_1$,label.side=right}{pone,v12}
      \fmf{plain}{ver,v2}
      \fmf{plain}{v21,v2}
      \fmf{fermion,label=$p'$,label.side=right}{v21,ptwo}
      \fmfright{pone,ptwo}
      \fmffreeze
      \fmfshift{(-.8cm,1cm)}{tip2}
    \fmf{photon,label=$k'$,label.side=right}{v12,tip2}
     \fmfshift{(1.7cm,0cm)}{tip1}
    \fmf{photon,label=$k$,label.side=right}{v1,tip1}
    \fmfdot{v1,ver,v12}
    \end{fmfgraph*}}
\end{center}
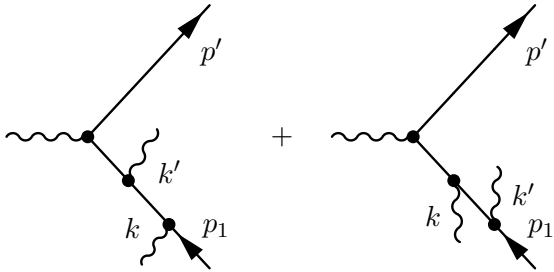
\caption{Collinear emission and absorption at lowest order.}}} In
both of these cases, the initial momentum $p$ is distributed between
the electron and photon, so $p=p_1+k$. We stress that $p_1$ and $k$
are on-shell but  $p$ is not. Clearly, collinear divergences will
arise if $k$ or $k'$ are parallel to $p_1$,

The unusual thing about this process is that to contribute to the
cross-section at order $e^4$, we need these emitting and absorbing
processes (which are already at order $e^3$) to interfere with a
process of order $e$. Following~\cite{Lee:1964is}, we consider the
\emph{disconnected} process shown in Figure~9 where the in-coming
electron is accompanied by a collinear photon that does \emph{not}
interact with it.

\FIGURE{\mbox{\begin{fmfgraph*}(30,35)\fmfstraight
      \fmfleft{source}
      \fmftopn{p}{1}
      \fmfbottomn{q}{1}
      \fmf{photon,l.d=1cm}{source,ver}
      \fmf{plain}{v1,ver}
      \fmf{fermion,label=$p_1$,label.side=right}{pone,v1}
      \fmf{plain}{ver,v2}
      \fmf{fermion,label=$p'$,label.side=right}{v2,ptwo}
      \fmfright{pone,ptwo}
      \fmf{photon}{p1,q1}
      \fmffreeze
      \fmfshift{(1.8cm,0)}{p1,q1}
      \fmfv{label=$k'$}{p1}
      \fmfv{label=$k$}{q1}
      \fmfdot{ver}
    \end{fmfgraph*}}
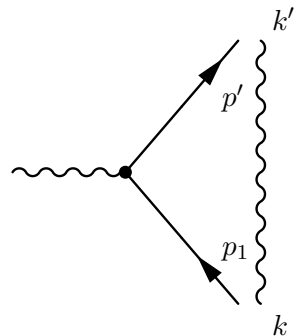
\caption{The basic disconnected process.}}

As the incoming photon is not observed, we should then integrate
over all allowed $k$ in the cross-section. The Feynman rule
associated with this disconnected process is
\begin{equation}\label{frule}
    -e\bar{u}'\gamma_0u_1\epsilon'\scdot\epsilon(2\pi)^32\omega'\delta^3(k'-k)\,.
\end{equation}
In standard discussions of the S-matrix, disconnected processes like
this are ignored as they describe the  no scattering situations
which can easily be distinguished experimentally. However, if we
have both initial and final degeneracies it is not possible to
distinguish them. As we will see they produce an important
interference with connected diagrams. The immediate effect, though,
of the disconnected photon line is to enforce $k=k'$ in the emitting
and absorbing processes described in Figure~8. This will then put
the internal electron line just before the scattering in the
diagrams of Figure~8 on-shell \emph{for all values of }$k'$. Clearly
this needs to be treated with great care.

How to deal with this is quite subtle and not fully addressed in the
literature. Indeed, Lee-Nauenberg handle the issue of vanishing
denominators by using a Hamiltonian approach in which,  for the
first diagram of Figure~8, the propagator between the emission and
absorption has the form
\begin{equation}\label{twos}
    \frac{\pslsh_1+\kslsh+m}{(p_1+k)^2-m^2+i(E_1+\omega)\alpha}\,,
\end{equation}
while the propagator before the scattering is
\begin{equation}\label{ones}
    \frac{\pslsh_1+\kslsh-\kslsh'+m}{(p_1+k-k')^2-m^2+2i(E_1+\omega-\omega')\alpha}\,.
\end{equation}
Here we see manifestly the double pole that arises  when $k=k'$ and
$p_1$ is on-shell. Also note the factor of 2 in the denominator
of~(\ref{ones}).  By retaining the imaginary parts, a vanishing
denominator is avoided at the expense of complex parameters. They
then extract the real part of the resulting cross-section. This
discussion is based on results from \lq\lq old fashioned"
perturbation theory, and does not make clear the need for a mass
resummation. We will take a more direct approach to these issues
here.

To proceed, we identify the S-matrix for the emission and absorption
processes  as
\begin{equation}\label{s8}
    ie^3\frac{\bar{u}'\gamma_0(\poneslsh+\kslsh-\kslsh')}{2p_1\scdot(k-k')-2k\scdot
    k'}\left\{\frac{\eslsh'(\poneslsh+\kslsh)\eslsh}{2p_1\scdot k}-\frac{\eslsh(
    \poneslsh+\kslsh)\eslsh'}{2p_1\scdot k'}\right\}u_1\,,
\end{equation}
where we recall that $u_1=u(p_1)$ and mass terms in the numerator
have been dropped. As discussed above, this will interfere with the
disconnected diagram and force $k=k'$. We  need to isolate the
collinear singularities from those that simply arise from this
identification of momenta. To this end, we need to identify those
terms in the numerator that vanish as $k\to k'$. As such, it is
helpful to write the S-matrix as the sum of two terms:
\begin{equation}\label{s9}
    ie^3\frac{\bar{u}'\gamma_0\poneslsh}{4(p_1\scdot(k-k'))(p_1\scdot
k)(p_1\scdot k')}\left\{(p_1\scdot
k')\eslsh'(\poneslsh+\kslsh)\eslsh-(p_1\scdot k)\eslsh(\poneslsh-
\kslsh')\eslsh'\right\}u_1
\end{equation}
and
\begin{equation}\label{s10}
    ie^3\frac{\bar{u}'\gamma_0(\kslsh-\kslsh')}{2(p_1\scdot (k-k'))(p_1\scdot
 k')}\left\{\eslsh'\kslsh'\eslsh'\right\}u_1\,.
\end{equation}
Here we have dropped the irrelevant sub-leading term  $2k\scdot
k'=-(k-k')^2$ in the denominator and in (\ref{s10}) we have, with an
eye to the interference with the disconnected diagram,  set $k=k'$
in the final braces as the numerator already contains a factor
of~$k-k'$.

The numerator in (\ref{s9}) can be written as ($ie^3$ times):
\begin{eqnarray}
&&-4(\epsilon'\scdot \epsilon')(p_1\scdot k')(p_1\scdot
k)\bar{u}'\gamma_0u_1-4 p_1\scdot (k-k')(p_1\scdot
\epsilon)(p_1\scdot \epsilon')\bar{u}'\gamma_0u_1\nonumber\\
&&+2p_1\scdot (k-k')p_1\scdot
(\epsilon'-\epsilon)\bar{u}'\gamma_0\eslsh\kslsh' u_1+2p_1\scdot
(k-k')(p_1\scdot \epsilon)\bar{u}'\gamma_0(\eslsh-\eslsh')\kslsh'
u_1\\\label{s11} &&+2(p_1\scdot k')p_1\scdot
(\epsilon'-\epsilon)\bar{u}'\gamma_0\eslsh(\kslsh'-\kslsh)
u_1+2(p_1\scdot k')(p_1\scdot
\epsilon)\bar{u}'\gamma_0(\eslsh-\eslsh')(\kslsh'-\kslsh)
u_1\,.\nonumber
\end{eqnarray}
The final four terms are now manifestly of order $(k-k')^2$  and can
be disregarded. The first term does not vanish as $k\to k'$ and
represents a double pole that arises in that limit. The second term
has no double pole and will contribute to the collinear structure of
this process.

The upshot of this is that we can write (\ref{s9}) in the limit as
$k\to k'$ as
\begin{equation}\label{s12}
    -ie^3\frac{2(\epsilon'\scdot
    \epsilon')}{p_1^2-m^2}\bar{u}'\gamma_0u_1-ie^3\frac{(p_1\scdot
\epsilon')(p_1\scdot \epsilon')}{(p_1\scdot k')(p_1\scdot
k')}\bar{u}'\gamma_0u_1\,,
\end{equation}
plus finite contributions. The first term here has been written in a
form that makes clear that it should be interpreted as a mass shift.
That such a mass shift should arise in this analysis should come as
no surprise since it is precisely these absorption/emission effects
with a background that lead to similar divergences when electrons
are in an intense laser background~\cite{kib} or at finite
temperature~\cite{DJ}.  The method for dealing with these effects
through a mass resummation is also now  understood~\cite{BP}. We
will assume that such a resummation has taken place and focus on the
collinear divergence that arise from the second term in~(\ref{s12}).

After multiplying by $-2ie\bar{u_1}\gamma_0u'$ (recall that the
factor of two comes from the cross terms that produce the
interference), summing over the polarisation $\lambda$ and
integrating over $k$, the second term in (\ref{s12}) yields the
following contribution to the cross-section
\begin{equation}\label{s13}
    -2e^4\frac{(p_1\scdot
\epsilon')(p_1\scdot \epsilon')}{(p_1\scdot k')(p_1\scdot
k')}|\bar{u}'\gamma_0u_1|^2=-2e^4\frac{(p_1\scdot
\epsilon')(p_1\scdot \epsilon')}{(p_1\scdot k')(p_1\scdot
k')}|\bar{u}'\gamma_0u|^2\frac{E_1}{E}\,,
\end{equation}
where we have used the collinear simplification (\ref{coul11}) and
dropped collinear finite terms. If we now sum over polarisations we
finally see that (\ref{s9}) contributes the collinear divergence
\begin{equation}\label{s14}
    -2e^4|\bar{u}'\gamma_0u|^2\frac{2E^2_1}{(p_1\scdot k')E\omega'}\,.
\end{equation}

The other part of this emmision/absorption process, given by
(\ref{s10}), contributes to the interference the term
\begin{equation}\label{s15}
    -e^4\frac{(\epsilon'\scdot
    \epsilon')}{p_1\scdot (k-k')(p_1\scdot
 k')}\mathrm{tr}\,(\tppslsh\,\kslsh\,\kslsh'\,\pslsh_1)\,.
\end{equation}
The trace can be readily evaluated to give
\begin{equation}\label{s16}
\mathrm{tr}\,(\tppslsh\,\kslsh\,\kslsh'\,\pslsh_1)=-p_1\scdot(k-k')|\bar{u}'\gamma_0u|^2\frac{\omega'}{E}
+\mathrm{collinear\ finite\ terms}\,.
\end{equation}
Hence we see that, after summing over polarisations, (\ref{s10})
contributes the collinear divergence
\begin{equation}\label{s17}
    -2e^4|\bar{u}'\gamma_0u|^2\frac{\omega'^2}{(p_1\scdot k')E\omega'}\,.
\end{equation}
Combining (\ref{s14}) and (\ref{s17}), and inserting the appropriate
energy weighting, we see that the interference contribution to the
collinear divergence is
\begin{equation}\label{s18}
P_{1,1}^{\mathrm{c}}(k',p_1):=-2e^4|\bar{u}'\gamma_0u|^2\frac{2E^2_1+\omega'^2}{(p_1\scdot
k')E\omega'}\frac{E}{E_1}\,.
\end{equation}
Here the Lee-Nauenberg probability $P_{1,1}^{\mathrm{c}}(k',p_1)$
describes the \emph{connected} interference contribution that arises
when there is one in-coming and one out-going collinear photon. (The
notation will be explained in full in the next section.) It is clear
that this does not cancel (\ref{s7}), so we need to look for another
degenerate process.
\FIGURE{\mbox{\begin{fmfgraph*}(30,35)\fmfstraight
      \fmfleft{source}
      \fmfbottom{tip1}
      \fmftopn{p}{1}
      \fmfbottomn{q}{1}
      \fmf{photon,l.d=1cm}{source,ver}
      \fmf{plain}{v1,ver}
      \fmf{fermion,label=$p_2$,label.side=right}{pone,v1}
      \fmf{plain}{ver,v2}
      \fmf{fermion,label=$p'$,label.side=right}{v2,ptwo}
      \fmfright{pone,ptwo}
      \fmf{photon}{p1,q1}
      \fmffreeze
     \fmfshift{(1.0cm,0)}{tip1}
    \fmf{photon,label=$k_2$,label.side=right}{v1,tip1}
      \fmffreeze
      \fmfshift{(1.8cm,0)}{p1,q1}
      \fmfv{label=$k'$}{p1}
      \fmfv{label=$k_1$}{q1}
      \fmfdot{v1,ver}
    \end{fmfgraph*}}
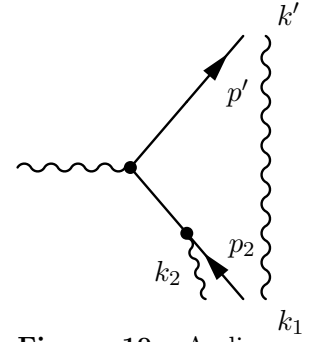
\caption{A disconnected process with two in-coming and one out-going
collinear photon. }} The only other collinearly divergent  process
that has a single photon in its final state is given in Figure~10
and arises when there are two in-coming photons, one of which gets
absorbed by the in-coming electron.
 The S-matrix for this is given by
\begin{equation}\label{cons}
    \frac{ie^2}{2p_2\scdot
    k_2}\bar{u}'\gamma_0(\ptwoslsh+\kslsh_2)\eslsh_2u_2\,\epsilon'
    \scdot\epsilon_1(2\pi)^32\omega'\delta^3(k'-k_1)\,,
\end{equation}
where now $u_2=u(p_2)$ and $p_2$ is on-shell.

This disconnected process can contribute to the cross-section at
order $e^4$ in two possible ways. If we represent the contraction of
photon lines by dashed lines, then  either the disconnected photon
lines are contracted together to yield a disconnected contribution
(as described in Figure~11(a)), or one gets the connected
contribution described in Figure~11(b). \FIGURE{
\begin{minipage}[t]{5cm}
\begin{fmfgraph*}(50,25)\fmfstraight
      \fmfleft{s1}
      \fmfright{s2}
      \fmfbottomn{q}{6}
      \fmftopn{p}{6}
      \fmffreeze
      \fmf{photon,l.d=1cm}{s1,ver1}
      \fmf{plain}{v12,ver1}
      \fmf{plain}{v11,ver1}
      \fmf{fermion}{q2,v11}
      \fmf{fermion}{v12,p2}
      \fmffreeze
      \fmf{photon,l.d=1cm}{s2,ver2}
      \fmf{plain}{v22,ver2}
      \fmf{plain}{v21,ver2}
      \fmf{fermion}{v21,q5}
      \fmf{fermion}{p5,v22}
      \fmffreeze
      \fmf{photon}{p3,q3}
      \fmf{photon}{p4,q4}
      \fmffreeze
    \fmf{photon}{v11,q1}
    \fmf{photon}{v21,q6}
      \fmffreeze
      \fmfshift{(1.4cm,0)}{q1}
      \fmfshift{(-1.4cm,0)}{q6}
     \fmf{dashes,left=.3}{p3,p4}
     \fmf{dashes,right=.3}{q3,q4}
     \fmf{dashes,right=.5}{q1,q6}
     \fmfdot{v11,ver1,v21,ver2}
    \end{fmfgraph*}\\\begin{center}
    (a)
    \end{center}
\end{minipage}
\qquad\begin{minipage}[t]{5cm}{\begin{fmfgraph*}(50,25)\fmfstraight
      \fmfleft{s1}
      \fmfright{s2}
      \fmfbottomn{q}{6}
      \fmftopn{p}{6}
      \fmffreeze
      \fmf{photon,l.d=1cm}{s1,ver1}
      \fmf{plain}{v12,ver1}
      \fmf{plain}{v11,ver1}
      \fmf{fermion}{q2,v11}
      \fmf{fermion}{v12,p2}
      \fmffreeze
      \fmf{photon,l.d=1cm}{s2,ver2}
      \fmf{plain}{v22,ver2}
      \fmf{plain}{v21,ver2}
      \fmf{fermion}{v21,q5}
      \fmf{fermion}{p5,v22}
      \fmffreeze
      \fmf{photon}{p3,q3}
      \fmf{photon}{p4,q4}
      \fmffreeze
    \fmf{photon}{v11,q1}
    \fmf{photon}{v21,q6}
      \fmffreeze
      \fmfshift{(1.4cm,0)}{q1}
      \fmfshift{(-1.4cm,0)}{q6}
     \fmf{dashes,left=.3}{p3,p4}
     \fmf{dashes,right=.3}{q3,q6}
     \fmf{dashes,right=.3}{q1,q4}
     \fmfdot{v11,ver1,v21,ver2}
    \end{fmfgraph*}}\\\begin{center}
    (b)
    \end{center} \end{minipage}
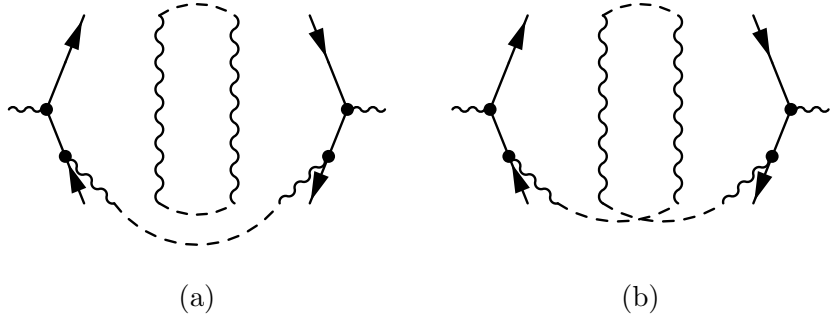
\caption{The two possible contractions of the photon lines. }}

The disconnected contraction is ignored by Lee-Nauenberg and we will
follow their lead on this and postpone  any discussion of it until
Section~6. The connected contraction is easy to calculate once it
has been understood that the \lq\lq loop" of photons can be simply
unwound to give a direct contraction of the absorbed photon (see
Appendix~A).

Indeed, the connected contraction is given by
\begin{equation}\label{cons2}
    -\frac{ie^2}{2p_2\scdot
    k_1}\bar{u}_2\eslsh_1(\ptwoslsh+\kslsh_1)\gamma_0u'\,\epsilon'
    \scdot\epsilon_2(2\pi)^32\omega'\delta^3(k'-k_2)\,.
\end{equation}
We now contract~(\ref{cons}) and~(\ref{cons2}) and integrate over
$k_1$ and $k_2$ and sum over two of the polarisations. This yields
the contribution
\begin{equation}\label{cons3}
    \frac{e^4}{(2p_2\scdot k')^2}\, \mathrm{tr}\,\left(\tppslsh(2p_2\scdot \epsilon'+
    \kslsh'\eslsh')\pslsh_2(2 p_2\scdot \epsilon'+\eslsh'\kslsh)\right)
\end{equation}
where now $p_2=p-2k'$ is on-shell and we still have the final
polarisation sum to perform. Note that this is precisely the
contribution that would have been obtained from the process of
absorbing a collinear photon of momentum $k'$ by an in-coming
electron with momentum $p_2$.

Proceeding as before, we readily arrive at the connected
contribution to the cross-section
\begin{equation}\label{cons4}
P_{2,1}^{\mathrm{c}}(k',p_2):=e^4|\bar{u}'\gamma_0
u|^2\frac{2E_1E_2+\omega'^2}{(p_2\scdot k')E\omega'}\frac{E}{E_2}\,,
\end{equation}
where the final term is the energy weighting for this process, we
recall that $E_1=E-\omega'$ and so define $E_2=E-2\omega'$.

We have so obtained the  relevant collinear  parts of three
degenerate contributions to the process of emitting a collinear jet
with one photon in the final state: $P_{0,1}$,
$P_{1,1}^{\mathrm{c}}$ and $P_{2,1}^{\mathrm{c}}$. Each of these
terms are collinearly divergent when the out-going photon's momenta
$k'$ is integrated over a small cone. However, as we shall now show,
the sum is finite.

To compare the three expressions (\ref{s7}), (\ref{s18}) and
(\ref{cons4}) we need to understand the relation between the terms
in the denominator. In Appendix~B we show that we may make the
replacement
\begin{equation}\label{cons5}
    \frac1{p\scdot k'}=\frac1{p_1\scdot
    k'}\frac{E}{E_1}=\frac1{p_2\scdot k'}\frac{E}{E_2}\,.
\end{equation}
Hence,
\begin{equation}\label{cons6}
    P_{0,1}+P_{1,1}^{\mathrm{c}}+P_{2,1}^{\mathrm{c}}=2e^4|\bar{u}'\gamma_0
u|^2\frac{ E_1(E+E_2)-2E_1^2}{(p\scdot k')E\omega'}\,.
\end{equation}
The cancellation now follows from the simple identity that
$E+E_2=2E_1$.

With this result  we have reached the end of Lee-Nauenberg's
argument. The distinctive feature of this work, which is not
stressed in the literature, is the need to incorporate disconnected
diagrams. In the introduction to~\cite{Lee:1964is}, Lee and
Nauenberg conclude with the statement that, \lq\lq \emph{Throughout
this paper the question of convergence of the power series is not
discussed}\rq\rq. We take this to mean that they recognised the need
to incorporate degeneracies where more photons are emitted at the
same order in the coupling, i.e., more disconnected photons. We will
now turn to this.

\end{fmffile}


\section{Filling out the jet}

\begin{fmffile}{filling}

In the previous section we have seen how collinear divergences
cancel in the jet process where a collinear photon is emitted from
the in-coming electron. This was achieved by considering all
processes degenerate with this single photon final state
configuration. The surprising thing about the mechanism for the
cancellation was the need to include in the cross-section the
connected interference contribution from disconnected S-matrix
elements. The analysis was  performed at order $e^4$ in the coupling
and is a striking demonstration of the Lee-Nauenberg theorem.

However, having opened the Pandora's box of disconnected diagrams,
we see that this is not the end of the story even for this simple
process. At the same order in the coupling we can also consider the
degenerate processes where there are two or more photons in the
final state. As an example, in Figure~\ref{fourjet} we see a four
photon final state process degenerate with the basic one photon
event described in Figure~6 and occurring at the \emph{same order in
perturbation theory}. Clearly, the out-going jet can be filled with
an arbitrary number of photons and still be degenerate to the
original process. We now need to understand how to deal with these
extra divergent processes. \FIGURE{\parbox{4cm}{
\begin{fmfgraph*}(30,35)
      \fmfstraight
      \fmfleft{source}
       \fmftopn{p}{4}
      \fmfbottomn{q}{4}
      \fmf{photon,l.d=2cm}{source,ver}
      \fmf{plain}{v1,ver}
      \fmf{fermion}{pone,v1}
      \fmf{plain}{ver,v2}
      \fmf{fermion}{v2,ptwo}
      \fmfright{pone,ptwo}
      \fmffreeze
     \fmfshift{(2.2cm,-1.7cm)}{p1}
    \fmf{photon}{v1,p1}
    \fmfdot{v1,ver}
    \fmf{photon}{p2,q2}
      \fmf{photon}{p3,q3}
      \fmf{photon}{p4,q4}
      \fmffreeze
      \fmfshift{(2.4cm,0)}{p2,q2}
      \fmfshift{(1.6cm,0)}{p3,q3}
      \fmfshift{(.2cm,0)}{p4,q4}
    \end{fmfgraph*}}\label{fourjet}
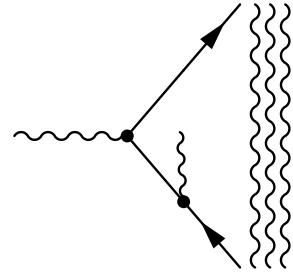
\caption{Filling the jet with four photons}}

So as not to get lost in the combinatorial details, it is best to
first consider the process which we denote
$P_{1,2}(k,\{k_a'\},p_{\mathrm{in}})$, i.e., the Lee-Nauenberg
probability corresponding to one in-coming photon with momentum $k$,
two out-going photons with momenta $k_1'$ and $k_2'$ and an
in-coming electron with momentum $p_{\mathrm{in}}$. These momenta
are not arbitrary but satisfy the constraints that $k_1'+k_2'=k'$
and $p_{\mathrm{in}}+k=p$. In order to extract the connected
component of $P_{1,2}$ we draw the skeleton diagram shown in
Figure~13(a) where the lower dashed line represents the single
in-state contraction and the two upper dashed lines the two
out-state contractions. This process comes with a symmetry factor of
$1/2!$ due to the two photon out-states.

\smallskip

\FIGURE{\begin{minipage}[t]{3.5cm}{\begin{fmfgraph*}(35,25)\fmfstraight
      \fmfleft{s1}
      \fmfright{s2}
      \fmfbottomn{q}{8}
      \fmftopn{p}{8}
      \fmffreeze
      \fmfshift{(.5cm,0)}{p2}
      \fmfshift{(.5cm,0)}{q2}
      \fmfshift{(-.5cm,0)}{p7}
      \fmfshift{(-.5cm,0)}{q7}
      \fmfshift{(.25cm,.2cm)}{p3}
      \fmfshift{(.25cm,-.2cm)}{q3}
      \fmfshift{(-.25cm,.2cm)}{p6}
      \fmfshift{(-.25cm,-.2cm)}{q6}
       \fmffreeze
      \fmf{photon,l.d=1cm}{s1,ver1}
      \fmf{plain}{v12,ver1}
      \fmf{plain}{v11,ver1}
      \fmf{fermion}{q2,v11}
      \fmf{fermion}{v12,p2}
      \fmffreeze
      \fmf{photon,l.d=1cm}{s2,ver2}
      \fmf{plain}{v22,ver2}
      \fmf{plain}{v21,ver2}
      \fmf{fermion}{v21,q7}
      \fmf{fermion}{p7,v22}
      \fmffreeze
    \fmf{photon}{v11,p1}
    \fmf{photon}{v21,p8}
      \fmffreeze
      \fmfshift{(1cm,-1.4cm)}{p1}
      \fmfshift{(-1cm,-1.4cm)}{p8}
     \fmf{dashes,left=.3}{p3,p6}
     \fmf{dashes,left=.3}{p4,p5}
     \fmf{dashes,right=.3}{q3,q6}
     \fmfdot{v11,ver1,v21,ver2}
    \end{fmfgraph*}}\\\begin{center}
    (a)
    \end{center} \end{minipage}
\qquad\begin{minipage}[t]{3.5cm}{\begin{fmfgraph*}(35,25)\fmfstraight
      \fmfleft{s1}
      \fmfright{s2}
      \fmfbottomn{q}{8}
      \fmftopn{p}{8}
      \fmffreeze
       \fmfshift{(.5cm,0)}{p2}
      \fmfshift{(.5cm,0)}{q2}
      \fmfshift{(-.5cm,0)}{p7}
      \fmfshift{(-.5cm,0)}{q7}
      \fmfshift{(.25cm,.2cm)}{p3}
      \fmfshift{(.25cm,-.2cm)}{q3}
      \fmfshift{(-.25cm,.2cm)}{p6}
      \fmfshift{(-.25cm,-.2cm)}{q6}
       \fmffreeze
      \fmf{photon,l.d=1cm}{s1,ver1}
      \fmf{plain}{v12,ver1}
      \fmf{plain}{v11,ver1}
      \fmf{fermion}{q2,v11}
      \fmf{fermion}{v12,p2}
      \fmffreeze
      \fmf{photon,l.d=1cm}{s2,ver2}
      \fmf{plain}{v22,ver2}
      \fmf{plain}{v21,ver2}
      \fmf{fermion}{v21,q7}
      \fmf{fermion}{p7,v22}
      \fmffreeze
    \fmf{photon}{v11,p3}
    \fmf{photon}{v21,p8}
      \fmffreeze
      \fmfshift{(1cm,-1.4cm)}{p1}
      \fmfshift{(-1cm,-1.4cm)}{p8}
     \fmf{dashes,left=.3}{p3,p6}
     \fmf{dashes,left=.3}{p4,p5}
     \fmf{dashes,right=.3}{q3,q6}
     \fmfdot{v11,ver1,v21,ver2}
    \end{fmfgraph*}}\\\begin{center}
    (b)
    \end{center} \end{minipage}
    \qquad\begin{minipage}[t]{3.5cm}{\begin{fmfgraph*}(35,25)\fmfstraight
      \fmfleft{s1}
      \fmfright{s2}
      \fmfbottomn{q}{8}
      \fmftopn{p}{8}
      \fmffreeze
       \fmfshift{(.5cm,0)}{p2}
      \fmfshift{(.5cm,0)}{q2}
      \fmfshift{(-.5cm,0)}{p7}
      \fmfshift{(-.5cm,0)}{q7}
      \fmfshift{(.25cm,.2cm)}{p3}
      \fmfshift{(.25cm,-.2cm)}{q3}
      \fmfshift{(-.25cm,.2cm)}{p6}
      \fmfshift{(-.25cm,-.2cm)}{q6}
       \fmffreeze
      \fmf{photon,l.d=1cm}{s1,ver1}
      \fmf{plain}{v12,ver1}
      \fmf{plain}{v11,ver1}
      \fmf{fermion}{q2,v11}
      \fmf{fermion}{v12,p2}
      \fmffreeze
      \fmf{photon,l.d=1cm}{s2,ver2}
      \fmf{plain}{v22,ver2}
      \fmf{plain}{v21,ver2}
      \fmf{fermion}{v21,q7}
      \fmf{fermion}{p7,v22}
      \fmffreeze
    \fmf{photon}{v11,p3}
    \fmf{photon}{p4,q3}
    \fmf{photon}{p6,q6}
    \fmf{photon,rubout}{v21,p5}
      \fmffreeze
      \fmfshift{(1cm,-1.4cm)}{p1}
      \fmfshift{(-1cm,-1.4cm)}{p8}
     \fmf{dashes,left=.3}{p3,p6}
     \fmf{dashes,left=.3}{p4,p5}
     \fmf{dashes,right=.3}{q3,q6}
     \fmfdot{v11,ver1,v21,ver2}
    \end{fmfgraph*}}\\\begin{center}
    (c)
    \end{center} \end{minipage}
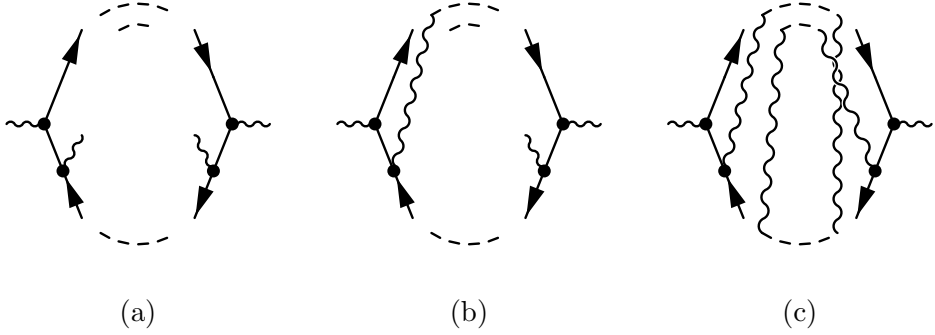
\caption{Building up the connected contribution to $P_{1,2}$. }} We
now need to connect the dashed lines by the photon lines in such a
way that we produce a connected diagram. Starting with the emitted
photon in Figure~13(b), we can connect it to one of the out-going
contraction lines in two possible ways. This figure has now a
symmetry factor of $2/2!=1$ reflecting the two possible
contractions. Once this line has been chosen, though, there is then
no freedom in completing  the rest of the contractions if we are
looking for a connected contribution. Thus we arrive at Figure~13(c)
with a symmetry factor of 1. The momentum delta functions now undo
the photon loop by setting $k=k_1'=k_2'=k'/2$ and we  arrive at an
expression equivalent to the original process $P_{0,1}$ but now with
momenta $k'/2$ for the out-going photon and
$p_{\frac12}=p-\frac12k'$ for the in-coming electron. Thus we have
shown that the Lee-Nauenberg probability $P_{1,2}$ has a connected
contribution, $P_{1,2}^{\mathrm{c}}$, to the cross-section where
\begin{equation}\label{fill1}
    P_{1,2}^{\mathrm{c}}(k,\{k_a'\},\pin)=P_{0,1}(\tfrac12k',p_{\frac12})\,.
\end{equation}
This analysis can be readily extended to the process
$P_{n-1,n}(\{k_i\},\{k'_a\},\pin)$ where we have $n-1$ in-coming
photons, $k_i$ and $n$ out-going photons, $k'_a$. As before, these
momenta are constrained so that:
\begin{equation}\label{fill2}
    \sum_{a=1}^n k'_a=k'\qquad \mathrm{and} \qquad \pin+ \sum_{i=1}^{n-1}
    k'_i=p\,.
\end{equation}
To represent this as in Figure~13, we would need $n$ dashed lines on
top and $n-1$ on the bottom. The symmetry factor for this would be
$1/n!(n-1)!$. Connecting the emitted photon with an upper line, as
in Figure~13(b), can be done in $n$ ways. Keeping the process
connected then restricts the next choice to $n-1$ possibilities.
Repeating this argument, we have $n!(n-1)!$ contractions leading to
the connected process and hence a final symmetry factor of, again,
1. The connected contribution from this is then
\begin{equation}\label{fill3}
    P_{n-1,n}^{\mathrm{c}}(\{k_i\},\{k'_a\},\pin)=P_{0,1}(\tfrac1nk',p_{\frac{n-1}{n}})\,.
\end{equation}
In a similar  way, we can consider the Lee-Nauenberg probabilities
with $n$ in-coming and $n$ out-going photons, $P_{n,n}$ and finally
the case with $n+1$ in-coming and $n$ out-going photons $P_{n+1,n}$.
What we find is that these have connected contributions related to,
respectively,  $P_{1,1}^{\mathrm{c}}$ and $P_{2,1}^{\mathrm{c}}$
which were introduced in Section~3. To be precise
\begin{equation}\label{fill4}
   P_{n,n}=  P_{n,n}(\{k_i\},\{k'_a\},\pin)\,,
\end{equation}
where
\begin{equation}\label{fill5}
    \sum_{a=1}^n k'_a=k'\qquad \mathrm{and} \qquad \pin+ \sum_{i=1}^{n}
    k'_i=p\,,
\end{equation}
and
\begin{equation}\label{fill6}
   P_{n+1,n}=  P_{n+1,n}(\{k_i\},\{k'_a\},\pin)\,,
\end{equation}
where now
\begin{equation}\label{fill7}
    \sum_{a=1}^n k'_a=k'\qquad \mathrm{and} \qquad \pin+ \sum_{i=1}^{n+1}
    k'_i=p\,.
\end{equation}
The connected contributions are then:
\begin{equation}\label{fill8}
    P_{n,n}^{\mathrm{c}}(\{k_i\},\{k'_a\},\pin)=P_{1,1}^{\mathrm{c}}(\tfrac1nk',p_1)\,,
\end{equation}
and
\begin{equation}\label{fill9}
    P_{n+1,n}^{\mathrm{c}}(\{k_i\},\{k'_a\},\pin)=P_{2,1}^{\mathrm{c}}(\tfrac1nk',p_{\frac{n+1}{n}})\,.
\end{equation}
There are no other degenerate processes at this order in
perturbation theory.

Having reduced the general connected Lee-Nauenberg probabilities to
re-scaled versions of the single photon probabilities calculated
earlier, it is now straightforward to evaluate them and we find:
\begin{eqnarray}
  P^{\mathrm{c}}_{n-1,n} &=& \frac{e^4|\bar{u}'\gamma_0u|^2}{(p\scdot k')E \omega'}(2n^2E_1E_{\frac{n-1}n}+\omega'^2)\,;\\
  P^{\mathrm{c}}_{n,n} &=& -2\frac{e^4|\bar{u}'\gamma_0u|^2}{(p\scdot k')E \omega'}(2n^2E_1^2+\omega'^2)\,;  \\
  P^{\mathrm{c}}_{n+1,n} &=& \frac{e^4|\bar{u}'\gamma_0u|^2}{(p\scdot k')E
  \omega'}(2n^2E_1E_{\frac{n+1}n}+\omega'^2)\,,
\end{eqnarray}
where now we define $E_m=E-m\omega'$.  Note that
\begin{equation}\label{fill10}
    P^{\mathrm{c}}_{n-1,n}+P^{\mathrm{c}}_{n,n}+P^{\mathrm{c}}_{n+1,n}=
    \frac{e^4|\bar{u}'\gamma_0u|^2}{(p\scdot k')E
    \omega'}2n^2(E_1E_{\frac{n-1}n}+E_1E_{\frac{n+1}n}-2E_1^2)\,.
\end{equation}
Hence, using the identity that
$E_{\frac{n-1}n}+E_{\frac{n+1}n}=2E_1$, we see that the sum is zero.
\end{fmffile}

\section{Soft divergences revisited}
\begin{fmffile}{soft}
Having seen the essential role played by disconnected processes that
produce a connected contribution to the cross-section, we can now
revisit the problem left unanswered in Section~2 and investigate the
cancellation of soft infrared divergences within the Lee-Nauenberg
framework.

Recall that, from equation~(\ref{coul4}), we have seen that the
infrared divergent content of  Coulombic scattering accompanied by
the emission of soft photons can be expressed as
\begin{equation}\label{bn1}
    \frac{d\sigma_0}{d\Omega}\frac{e^2}{4\pi^2}\left(\!\ln\left(\!
\frac{\!Q^2}{m^2}\right)-1\right)\frac1{\hat{\varepsilon}_{_\mathrm{IR}}}(-1+1)=0\,.
\end{equation}
Here the $-1$ term  comes from the virtual contribution and the $+1$
from the real emission. Hence the content of the Bloch-Nordsieck
approach to soft infrared cancellation can be succinctly summarised
by the formula
\begin{equation}\label{bn1.1}
    \frac1{\hat{\varepsilon}_{_\mathrm{IR}}}(-1+1)=0\,.
\end{equation}
\FIGURE{\parbox{30mm}{\begin{fmfgraph*}(30,35)
    \fmfleft{source}
     \fmftop{tip1}
      \fmf{photon,l.d=2cm}{source,ver}
      \fmf{plain}{v1,ver}
      \fmf{fermion,label=$p$,label.side=right}{pone,v1}
      \fmf{plain}{ver,v2}
      \fmf{fermion,label=$p'$,label.side=right}{v2,ptwo}
      \fmfright{pone,ptwo}
      \fmffreeze
      \fmfshift{(.4cm,-1.8cm)}{tip1}
    \fmf{photon,label=$k$,label.side=left}{v2,tip1}
    \end{fmfgraph*}}
    \quad + \quad
\parbox{30mm}{\begin{fmfgraph*}(30,35)
      \fmfleft{source}
     \fmfbottom{tip1}
      \fmf{photon,l.d=2cm}{source,ver}
      \fmf{plain}{v1,ver}
      \fmf{fermion,label=$p$,label.side=right}{pone,v1}
      \fmf{plain}{ver,v2}
      \fmf{fermion,label=$p'$,label.side=right}{v2,ptwo}
      \fmfright{pone,ptwo}
      \fmffreeze
      \fmfshift{(.7cm,0)}{tip1}
    \fmf{photon,label=$k$,label.side=right}{v1,tip1}
    \end{fmfgraph*}}
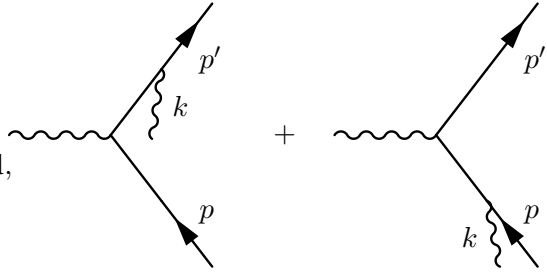
    \caption{Soft absorption to the out-going and in-coming electron. }}

In the Lee-Nauenberg approach we must also consider initial
degeneracies and the associated absorption of soft photons as shown
in Figure~14. As we have seen in other examples of the Lee-Nauenberg
method, these are taken to contribute is exactly the same way as the
emission process. That is, although it is possible to have a
different initial resolution which will change finite terms, the
essential infrared divergence will be the same if the ensemble of
soft photons has the same distribution for both in and out states.
Hence, including these initial processes, we find that the soft
infrared content of the cross-section becomes
\begin{equation}\label{bn2}
    \frac1{\hat{\varepsilon}_{_\mathrm{IR}}}(-1+1+1)\,.
\end{equation}
Clearly this is no longer zero and we have an infrared divergent
cross-section.

Of course, we should not stop here. What we have learnt from our
analysis of Lee-Nauenberg is that we must include \emph{all}
degenerate processes and hence need to consider the processes
described in Figure~15. Including soft photons compounds the number
of degenerate processes as it is now not possible to identify which
electron an undetectable soft photon was emitted from or absorbed
by.
 \FIGURE{\parbox{12cm}{
    \parbox{30mm}{\begin{fmfgraph*}(30,35)
    \fmfleft{source}
     \fmftop{tip1,tip2}
      \fmf{photon,l.d=2cm}{source,ver}
      \fmf{plain}{v1,ver}
      \fmf{plain}{v12,v1}
      \fmf{fermion,label=$p$,label.side=right}{pone,v12}
      \fmf{plain}{ver,v2}
      \fmf{plain}{v21,v2}
      \fmf{fermion,label=$p'$,label.side=right}{v21,ptwo}
      \fmfright{pone,ptwo}
      \fmffreeze
      \fmfshift{(2.2cm,.3cm)}{tip1}
    \fmf{photon,label=$k'$,label.side=left}{v21,tip1}
     \fmfshift{(-1.5cm,-1.5cm)}{tip2}
    \fmf{photon,label=$k$,label.side=left}{v2,tip2}
    \end{fmfgraph*}\\\begin{center}
    (a)
    \end{center}
    }\qquad
\parbox{30mm}{\begin{fmfgraph*}(30,35)
      \fmfleft{source}
     \fmftop{tip1,tip2}
      \fmf{photon,l.d=2cm}{source,ver}
      \fmf{plain}{v1,ver}
      \fmf{plain}{v12,v1}
      \fmf{fermion,label=$p$,label.side=right}{pone,v12}
      \fmf{plain}{ver,v2}
      \fmf{plain}{v21,v2}
      \fmf{fermion,label=$p'$,label.side=right}{v21,ptwo}
      \fmfright{pone,ptwo}
      \fmffreeze
      \fmfshift{(-1cm,-1cm)}{tip2}
    \fmf{photon,label=$k$,label.side=left}{v21,tip2}
     \fmfshift{(1.7cm,0cm)}{tip1}
    \fmf{photon,label=$k'$,label.side=left}{v2,tip1}
    \end{fmfgraph*}\begin{center}
    (b)
    \end{center}}
    \qquad\parbox{30mm}{\begin{fmfgraph*}(30,35)
    \fmfleft{source}
     \fmfbottom{tip2,tip1}
      \fmf{photon,l.d=2cm}{source,ver}
      \fmf{plain}{v1,ver}
      \fmf{plain}{v12,v1}
      \fmf{fermion,label=$p$,label.side=right}{pone,v12}
      \fmf{plain}{ver,v2}
      \fmf{plain}{v21,v2}
      \fmf{fermion,label=$p'$,label.side=right}{v21,ptwo}
      \fmfright{pone,ptwo}
      \fmffreeze
      \fmfshift{(-1.2cm,-.3cm)}{tip1}
    \fmf{photon,label=$k$,label.side=right}{v12,tip1}
     \fmfshift{(2cm,1.5cm)}{tip2}
    \fmf{photon,label=$k'$,label.side=right}{v1,tip2}
    \end{fmfgraph*}\begin{center}
    (c)
    \end{center}}\\
\parbox{30mm}{\begin{fmfgraph*}(30,35)
      \fmfleft{source}
     \fmfbottom{tip1,tip2}
      \fmf{photon,l.d=2cm}{source,ver}
      \fmf{plain}{v1,ver}
      \fmf{plain}{v12,v1}
      \fmf{fermion,label=$p$,label.side=right}{pone,v12}
      \fmf{plain}{ver,v2}
      \fmf{plain}{v21,v2}
      \fmf{fermion,label=$p'$,label.side=right}{v21,ptwo}
      \fmfright{pone,ptwo}
      \fmffreeze
      \fmfshift{(-.8cm,1cm)}{tip2}
    \fmf{photon,label=$k'$,label.side=right}{v12,tip2}
     \fmfshift{(1.7cm,0cm)}{tip1}
    \fmf{photon,label=$k$,label.side=right}{v1,tip1}
    \end{fmfgraph*}\begin{center}
    (d)
    \end{center}}\qquad\parbox{30mm}{\begin{fmfgraph*}(30,35)
    \fmfleft{source}
     \fmftop{tip1}
     \fmfbottom{tip2}
      \fmf{photon,l.d=2cm}{source,ver}
      \fmf{plain}{v1,ver}
      \fmf{fermion,label=$p$,label.side=right}{pone,v1}
      \fmf{plain}{ver,v2}
      \fmf{fermion,label=$p'$,label.side=right}{v2,ptwo}
      \fmfright{pone,ptwo}
      \fmffreeze
      \fmfshift{(-.5cm,0)}{tip1}
    \fmf{photon,label=$k'$,label.side=left}{v2,tip1}
    \fmfshift{(.7cm,0)}{tip1}
    \fmf{photon,label=$k$,label.side=right}{v1,tip2}
    \end{fmfgraph*}\begin{center}
    (e)
    \end{center}}
    \qquad
    \parbox{30mm}{\begin{fmfgraph*}(30,35)
    \fmfleft{source}
     \fmftop{tip1}
      \fmfbottom{tip2}
      \fmf{photon,l.d=2cm}{source,ver}
      \fmf{plain}{v1,ver}
      \fmf{fermion,label=$p$,label.side=right}{pone,v1}
      \fmf{plain}{ver,v2}
      \fmf{fermion,label=$p'$,label.side=right}{v2,ptwo}
      \fmfright{pone,ptwo}
      \fmffreeze
      \fmfshift{(1.1cm,-1.6cm)}{tip1}
    \fmf{photon,label=$k$,label.side=right}{v2,tip1}
    \fmfshift{(1.1cm,1.5cm)}{tip2}
    \fmf{photon,label=$k'$,label.side=left}{v1,tip2}
    \end{fmfgraph*}\begin{center}
    (f)
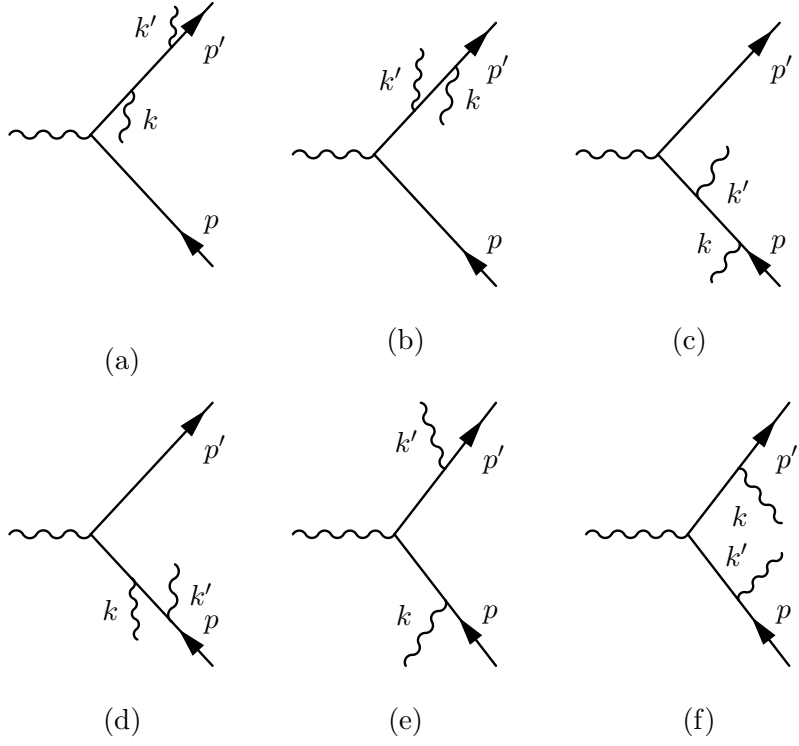
    \end{center}}}\caption{Soft emission and absorption process.}}

\noindent The S-matrix element corresponding to diagrams (a) and (b)
in Figure~15 is:
\begin{eqnarray}\label{bn3}
    &&(ie)^3\overline{u}'\eslsh'\frac{i}{\ppslsh+\kslsh'-m}
    \eslsh\frac{i}{\ppslsh+\kslsh'-\kslsh-m}\gamma_0u\nonumber\\&&\qquad+
(ie)^3\overline{u}'\eslsh\frac{i}{\ppslsh-\kslsh-m}
    \eslsh'\frac{i}{\ppslsh+\kslsh'-\kslsh-m}\gamma_0u\,.
\end{eqnarray}
Note that, due to the softness of the photons,  here we can take the
momentum of the electrons to be just $p$ and $p'$.   Just keeping
the leading order soft elements of this (so we assume that the soft
mass resummation has been carried out) we get the S-matrix
contribution
\begin{equation}\label{bn4}
    -ie^3\overline{u}'\gamma_0u \frac{p'\scdot\epsilon'}{p'\scdot k'}\frac{p'\scdot\epsilon}{p'\scdot
    k}\,.
\end{equation}
In a similar way, for the processes described in diagrams (c) and
(d) of Figure~15, we get
\begin{equation}\label{bn5}
    -ie^3\overline{u}'\gamma_0u \frac{p\scdot\epsilon'}{p\scdot k'}\frac{p\scdot\epsilon}{p\scdot
    k}\,.
\end{equation}
The final two  diagrams in Figure~15 are slightly different in
structure and give the contribution
\begin{equation}\label{bn6}
    ie^3\overline{u}'\gamma_0u\left( \frac{p'\scdot\epsilon'}{p'\scdot k'}\frac{p\scdot\epsilon}{p\scdot
    k}+\frac{p\scdot\epsilon'}{p\scdot k'}\frac{p'\scdot\epsilon}{p'\scdot
    k}\right)\,.
\end{equation}
Combining all of these absorption/emission processes we get the
S-matrix element
\begin{equation}\label{bn7}
    -ie^3\overline{u}'\gamma_0u\left[\left( \frac{p'\scdot\epsilon'}{p'\scdot k'}-\frac{p\scdot\epsilon'}{p\scdot
    k'}\right)\left(\frac{p'\scdot\epsilon}{p'\scdot k}-\frac{p\scdot\epsilon}{p\scdot
    k}\right)\right]\,.
\end{equation}
This is now contracted with the basic disconnected process described
in Figure~9 and integrated over the photon momenta, summed over the
photon polarisations and multiplied by 2 as it is a cross term in
the cross-section. This yields $-2$ times the emitted or absorbed
processes $R_s$ in (\ref{coul5}). Hence, by including these new
degenerate processes, we arrive at the soft infrared content
\begin{equation}\label{bn8}
    \frac1{\hat{\varepsilon}_{_\mathrm{IR}}}(-1+1+1-2)\,,
\end{equation}
to the cross-section. However, we still have a non-vanishing
infrared pole.

\bigskip

\FIGURE{\vspace*{4mm} \mbox{\begin{fmfgraph*}(30,30)\fmfstraight
      \fmfleft{source}
      \fmfbottom{tip1}
      \fmftopn{p}{1}
      \fmfbottomn{q}{1}
      \fmf{photon,l.d=1cm}{source,ver}
      \fmf{plain}{v1,ver}
      \fmf{fermion,label=$p$,label.side=right}{pone,v1}
      \fmf{plain}{ver,v2}
      \fmf{fermion,label=$p'$,label.side=right}{v2,ptwo}
      \fmfright{pone,ptwo}
      \fmf{photon}{p1,q1}
      \fmffreeze
     \fmfshift{(1.2cm,0)}{tip1}
    \fmf{photon,label=$k'_1$,label.side=left}{v1,tip1}
      \fmffreeze
     \fmfshift{(.5cm,1.5cm)}{tip1}
      \fmfshift{(2cm,0)}{p1,q1}
      \fmfv{label=$k'_2$}{p1}
      \fmfv{label=$k$}{q1}
      \fmfdot{v1,ver}
    \end{fmfgraph*}}
    \quad \quad\qquad
    \mbox{\begin{fmfgraph*}(30,30)\fmfstraight
      \fmfleft{source}
      \fmfbottom{tip1}
      \fmftopn{p}{1}
      \fmfbottomn{q}{1}
      \fmf{photon,l.d=1cm}{source,ver}
      \fmf{plain}{v1,ver}
      \fmf{fermion,label=$p$,label.side=right}{pone,v1}
      \fmf{plain}{ver,v2}
      \fmf{fermion,label=$p'$,label.side=right}{v2,ptwo}
      \fmfright{pone,ptwo}
      \fmf{photon}{p1,q1}
      \fmffreeze
     \fmfshift{(1.2cm,0)}{tip1}
    \fmf{photon,label=$k'_1$,label.side=left}{v2,tip1}
      \fmffreeze
     \fmfshift{(-0.2cm,3.2cm)}{tip1}
      \fmfshift{(2cm,0)}{p1,q1}
      \fmfv{label=$k'_2$}{p1}
      \fmfv{label=$k$}{q1}
      \fmfdot{v2,ver}
    \end{fmfgraph*}}
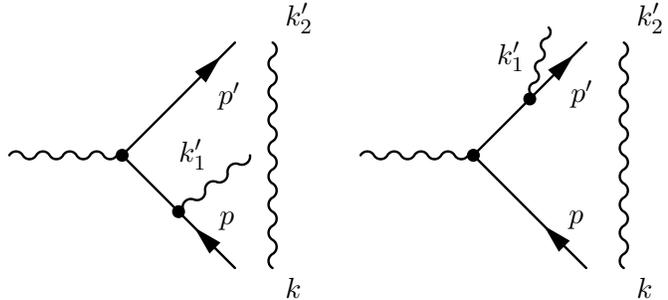
\caption{The disconnected processes with one in-coming and two
out-going soft photons. }}

To proceed, we need to include some more degenerate processes. For
example, we could include the process $P_{1,2}$ described in
Figure~16 (this is the soft version of Figure~13(a)). In this we
need to sum over all soft photons with the constraint that $0\le
k_1'+k_2'\le\Delta$. Following the process described in Figure~13
for extracting the connected contribution to the cross-section, we
will arrive at the process described in Figure~2 (with $k'$ replaced
by $k_1'$) with the constraint that $k_1'=k_2'=k$. Hence the
contribution to the cross-section will be as in
equation~(\ref{coul5}) but with the integration range up to
$\tfrac12\Delta$. However, as we have already discussed, the
infrared divergence is insensitive to the value of the  resolution
and hence the degenerate process $P_{1,2}$ will contribute the same
as $P_{0,1}$ to the infrared pole in the cross-section and
thus~(\ref{bn8}) becomes
\begin{equation}\label{bn9}
    \frac1{\hat{\varepsilon}_{_\mathrm{IR}}}(-1+1+1-2+1)\,.
\end{equation}
At this stage we have recovered an infrared finite cross-section and
this is essentially the argument given in \cite{Muta:1981pe}. What
this seems to show is that the virtual divergence, $P_{0,0}$ is
cancelled by the emission $P_{0,1}$ via the Bloch-Nordsieck
mechanism, and then the absorption process $P_{1,0}$ is cancelled by
the mixed processes $P^{\mathrm{c}}_{1,1}$ and
$P^{\mathrm{c}}_{1,2}$. That is,
\begin{equation}\label{bn10}
   0=
   -1+1+1-2+1=\underbrace{-1+1}_{\mathrm{Bloch-Nordsieck}}+\underbrace{1-2+1}_{\mathrm{Lee-Nauenberg}}\,.
\end{equation}
This result seems to reconcile the Bloch-Nordsieck and the
Lee-Nauenberg approaches to  soft infrared divergences. However, we
have so far only included one disconnected photon and, even at this
order in perturbation theory, we can and indeed must include an
arbitrary number of such photons.

\end{fmffile}
\section{The question of convergence}
We have seen that, in applying the Lee-Nauenberg theorem to quantum
field theory, what one does in practice is to include enough
degenerate states to achieve infrared finiteness for the
cross-section. For the two jet process discussed in Section~2, the
\emph{only} relevant degenerate states at order $e^4$ are those
where a photon is emitted from the out-going electron, or absorbed
by the in-coming electron. There are \emph{no} disconnected
semi-hard  contributions. This would also be the case in any process
without massless initial charged states.

 For the three jet event
discussed in Section~3, we have seen that disconnected contributions
are essential for the cancellation observed in (\ref{cons6}). We
have also just demonstrated in (\ref{bn9}) that this is the case for
soft cancellation if we go beyond Bloch-Nordsieck (which we have
seen in Section~2 is essential in order to capture the full
soft-collinear structure of the theory).

However, why should we stop at these levels of degeneracies? The
Lee-Nauenberg theorem requires us to sum over \emph{all}
degeneracies. In the three jet system we went beyond the discussion
given by Lee and Nauenberg and saw in (\ref{fill10}) that the
Lee-Nauenberg probabilities could be grouped in such a way that the
cancellation  seems to work for the jet full of disconnected
photons. A similar argument can be developed for the soft structures
discussed in the last section. Indeed, for soft momenta the
construction is greatly simplified since in, for example,
(\ref{fill3}), (\ref{fill8}) and (\ref{fill9}), the in-coming
electron always has momentum $p$. The upshot of this is that, as far
as the infrared pole is concerned, we have the identities:
\begin{equation}\label{con1}
P_{n-1,n}^{\mathrm{c}}=P_{0,1}\qquad
P_{n,n}^{\mathrm{c}}=P_{1,1}^{\mathrm{c}}\quad(n>0)\quad
\mathrm{and} \quad P_{n+1,n}^{\mathrm{c}}=P_{1,0}\,,
\end{equation}
along with the relations derived in Section~5 that
$P_{0,1}=-P_{0,0}$, $P_{1,1}^{\mathrm{c}}=-(P_{0,1}+P_{1,0})$ and
$P_{0,1}=P_{1,0}$.

The generalisation of~(\ref{bn9}) is then the vanishing of the
quantity
\begin{equation}\label{con2}
  P_{0,0}+P_{0,1}+\sum_{n=1}^\infty(P_{n,n-1}^{\mathrm{c}}+
  P_{n,n}^{\mathrm{c}}+P_{n,n+1}^{\mathrm{c}})\,.
\end{equation}
Now, however, we need to enquire as to the robustness of this
result. That is, we need to ensure that the series involved are
converging. Clearly they are not! This is most strikingly see in
(\ref{con2}), where we can simply absorb the $P_{0,1}$ term into the
sum to get the non-vanishing result:
\begin{equation}\label{con3}
  P_{0,0}+\sum_{n=1}^\infty(P_{n,n-1}^{\mathrm{c}}+
  P_{n,n}^{\mathrm{c}}+P_{n-1,n}^{\mathrm{c}})=P_{0,0}\,.
\end{equation}
In fact, this ordering is quite attractive from a coherent state
approach or also from the dressing
description~\cite{Lavelle:1995ty,Usannals} of gauge invariant
charged particles. This is because the virtual term $P_{0,0}$ is
already infrared finite in both of those approaches. This would then
show that the infrared finiteness is not spoilt by any undetectable
soft process. However, expressions like (\ref{con2}) or (\ref{con3})
cannot be taken seriously as it is clear that the sum is  ill
defined. Similar conclusions can be reached about the three jet
process. A more striking example of how the lack of convergence can
lead to unacceptable  results is presented in Appendix~C.

\section{Discussion and open problems}
We have seen that there are surprising aspects to the application of
the Lee-Nauenberg theorem in gauge theories. In particular, one
needs to include disconnected diagrams since they can produce
connected, interference contributions to cross-sections. These only
arise when there are both initial and final state degeneracies.
However, they are essential for all soft processes and for collinear
multi-jet processes where one of the jets is in the direction of the
incoming beam. The apparent cancellations which we have seen for
arbitrary numbers of photons are quite remarkable. However, we have
also seen that the cancellation is not robust as the expansion in
the number of photons is not convergent. We have also demonstrated
that there is a mass shift caused by these diagrams which implies
the need for a mass resummation in any phenomenological application.

One of the attractive aspects of these calculations is that one can
admit the physically intuitive picture that there can be
unobservable particles in both the final \emph{and} initial states.
The unattractive side to this is the requirement that the degenerate
initial and final states have the same ensemble distribution in
momentum space. This involves some fine-tuning of how the initial
state is prepared which has led to some
debate~\cite{Taylor:1993bz,Contopanagos:1996qx} on the preparation
of experimental initial states. Reservations on this are also
apparent in p.~1554 of~\cite{Lee:1964is} and, more recently, in
Weinberg's remarks on p.~552 of~\cite{Weinberg} \emph{\lq\lq The sum
over initial states is more problematic. Presumably one may argue
that truly massless particles are always produced as jets
accompanied by an ensemble of soft quanta that is uniform within
some volume of momentum space. However, to the best of my knowledge
no one has given a complete demonstration that the sums of
transition rates that are free of infrared divergences are the only
ones that are experimentally measurable."}

Another consequence of initial sums is the need for disconnected
processes. These in turn produce disconnected contributions to the
cross-section which are usually dropped in the literature. A naive
treatment of these terms introduces a factor of $\delta^{(3)}(0)$
which should be interpreted as a volume. These ill-defined terms
arise because we are using plane wave states. A more careful
treatment would use wave-packets concentrated around the beam, so
that the beam volume would replace this singularity. Such a
description has not yet been constructed.

Although we feel that our analysis has clarified important aspects
of the Lee-Nauenberg theorem, we have also seen that, as it stands,
it cannot be applied to any soft process or to any collinear process
with both initial and final state degeneracies which allows
interference from disconnected diagrams. Given the significance of
this theorem, we feel that it is essential that the following open
questions are answered.
\begin{itemize}
    \item We have seen that the connected interference terms do not
    converge for arbitrarily many disconnected photons. A naive
    attempt at making it converge by using a coherent state
    normalisation results in a breakdown of the infrared
    cancellations. Can we find a set of well defined particle states with
    which the infrared divergences cancel in a well-defined
    fashion at the level of perturbative cross-sections?
    \item How can we consistently treat  disconnected diagrams? Is there a
    way to treat them such that the disconnected contributions can
    be factorised out of the cross-section and hence absorbed into a
    normalisation? This again requires that the series of
    Lee-Nauenberg probabilities converges for arbitrarily many
    disconnected photons.
    \item The need for summing over initial
    particles was first recognised as important with the birth of QCD. How then
    can this analysis be extended to the non-abelian theory? We
    note that it is essential here to include the soft and collinear
    effects of three and four gluon vertices.
\end{itemize}
Although infrared  safety allows us to sidestep many of these
questions, a deeper understanding of gauge theories will, we feel,
follow from answering them.

\acknowledgments We wish to thank especially Emili Bagan for
detailed discussions on many of the topics treated in this paper. We
also thank Tom Heinzl, Paul Jameson, Arsen Khvedelidze and Oliver
Schr\"oder for discussions.

\appendix

\section{Contracting photon lines}
\begin{fmffile}{appen}
In many places in this paper we have  \lq\lq unwound\rq\rq connected
lines of photons. Diagrammatically, this can be summarised as the
identity shown in Figure~17. What we want to do in this Appendix is
show how this unwinding process follows from the Feynman
rule~(\ref{frule}).

\medskip

 \FIGURE{
\begin{minipage}[t]{5cm}{\begin{fmfgraph*}(50,25)\fmfstraight
      \fmfleft{s1}
      \fmfright{s2}
      \fmfbottomn{q}{6}
      \fmftopn{p}{6}
      \fmffreeze
      \fmf{photon,l.d=1cm}{s1,ver1}
      \fmf{plain}{v12,ver1}
      \fmf{plain}{v11,ver1}
      \fmf{fermion}{q2,v11}
      \fmf{fermion}{v12,p2}
      \fmffreeze
      \fmf{photon,l.d=1cm}{s2,ver2}
      \fmf{plain}{v22,ver2}
      \fmf{plain}{v21,ver2}
      \fmf{fermion}{v21,q5}
      \fmf{fermion}{p5,v22}
      \fmffreeze
      \fmf{photon}{p3,q3}
      \fmf{photon}{p4,q4}
      \fmffreeze
    \fmf{photon}{v11,q1}
    \fmf{photon}{v21,q6}
      \fmffreeze
      \fmfshift{(1.4cm,0)}{q1}
      \fmfshift{(-1.4cm,0)}{q6}
     \fmfv{label=$p'$}{p2}
      \fmfv{label=$p_2$}{q2}
     \fmf{dashes,left=.3}{p3,p4}
     \fmf{dashes,right=.3}{q3,q6}
     \fmf{dashes,right=.3}{q1,q4}
     \fmfdot{v11,ver1,v21,ver2}
    \end{fmfgraph*}} \end{minipage}
\begin{minipage}[t]{1.5cm}
\vspace*{-1.25cm}
    \begin{center}
    $=$\end{center}

    \end{minipage}
\begin{minipage}[t]{5cm}
\begin{fmfgraph*}(50,25)\fmfstraight
      \fmfleft{s1}
      \fmfright{s2}
      \fmfbottomn{q}{6}
      \fmftopn{p}{6}
      \fmffreeze
      \fmf{photon,l.d=1cm}{s1,ver1}
      \fmf{plain}{v12,ver1}
      \fmf{plain}{v11,ver1}
      \fmf{fermion}{q2,v11}
      \fmf{fermion}{v12,p2}
      \fmffreeze
      \fmf{photon,l.d=1cm}{s2,ver2}
      \fmf{plain}{v22,ver2}
      \fmf{plain}{v21,ver2}
      \fmf{fermion}{v21,q5}
      \fmf{fermion}{p5,v22}
      \fmffreeze
      \fmffreeze
    \fmf{photon}{v11,q1}
    \fmf{photon}{v21,q6}
      \fmffreeze
      \fmfshift{(1.4cm,0)}{q1}
      \fmfshift{(-1.4cm,0)}{q6}
     \fmfv{label=$p'$}{p2}
     \fmfv{label=$p_2$}{q2}
     \fmf{dashes,right=.5}{q1,q6}
     \fmfdot{v11,ver1,v21,ver2}
    \end{fmfgraph*}
\end{minipage}
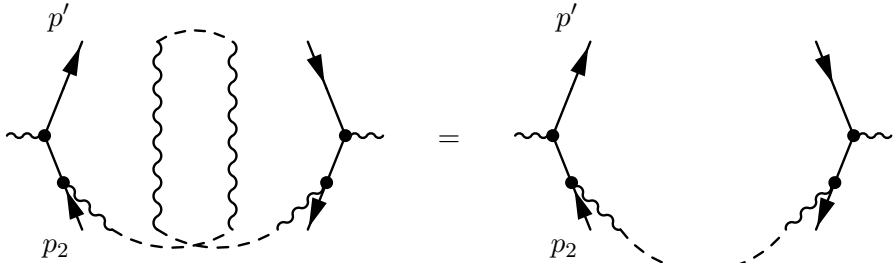
\caption{Unwinding connected loop contractions}}

To be concrete, using the momentum assignments in Figure~10, the
basic S-matrix element is
\begin{equation}\label{wind1}
    ie^2\frac{\bar{u}'\gamma_0(\pslsh_2+\kslsh_2)\eslsh_2u_2}{2p_2\scdot k_2}
    \epsilon'\scdot\epsilon_1(2\pi)^32\omega'\delta^3(k'-k_1)\,.
\end{equation}
In order to have a connected contribution to the cross-section, this
must be contracted with (see Figure~11(b))
\begin{equation}\label{wind2}
    -ie^2\frac{\bar{u}_2\eslsh_1(\pslsh_2+\kslsh_1)\gamma_0u'}{2p_2\scdot k_2}
    \epsilon'\scdot\epsilon_2(2\pi)^32\omega'\delta^3(k'-k_2)\,.
\end{equation}
We recall that in both (\ref{wind1}) and (\ref{wind2}) we are using
the notation that $\epsilon'=\epsilon(k',\lambda')$,
$\epsilon_1=\epsilon(k_1,\lambda_1)$ and
$\epsilon_2=\epsilon(k_2,\lambda_2)$.

Contracting these two terms together by integrating over the momenta
$k_1$ and $k_2$ which enforces all the momenta to be $k'$, we arrive
at an expression of the form
\begin{equation}\label{wind3}
    A_{\mu\nu}\sum_{\lambda',\lambda_1,\lambda_2}\epsilon^\mu(\lambda_2)\epsilon^\nu(\lambda_1)
    \epsilon_\rho(\lambda')\epsilon^\rho(\lambda_1)\epsilon_\tau(\lambda')\epsilon^\tau(\lambda_2)\,,
\end{equation}
where we have suppressed the common $k'$ dependence in the
polarisation tensors. Now summing over the polarisations $\lambda_1$
and $\lambda_2$, and making  repeated use of the identity
(\ref{coul9}), allows us to reduce this to
\begin{equation}\label{wind4}
    A_{\mu\nu}\sum_{\lambda'}\epsilon^\mu(k',\lambda')\epsilon^\nu(k',\lambda')\,.
\end{equation}
This is precisely the contribution that we would have written down
for the unwound process in  Figure~17, where there is just an
absorption of a photon of momentum $k'$ by the incoming electron
with momentum $p_2$.

\section{Some collinear approximations}
At various places in this paper we make approximations based on
collinearity. There are two classes of such approximations that we
want to discuss in some detail.
\begin{description}
    \item[Type 1] Terms of the form $p\scdot k'$ that occur in a
    numerator where $k'$ is not collinear with $p$, but is
    approximately collinear with $p'$. Here we want to
    understand how to approximate the scalar product by $p\scdot
    p'$.
    \item[Type 2] Terms of the form $p_n\scdot k'$ that occur in a
    denominator where $k'$ is approximately collinear with $p_n$, so the scalar product is very small.
    Here we want to understand how to approximate the  scalar product
    by $p\scdot k'$.
\end{description}

\paragraph{Type 1 approximations} Here we consider the non-vanishing $p\scdot k'$
\begin{equation}
p\scdot k'=E_p\omega'-\vert\underline{p}\vert \vert
\underline{k}'\vert \cos(\theta_{k'})\,,
\end{equation}
where $\theta_{k'}$ is the large angle between the momenta
$\underline{p}$ and $\underline{k}'$. Neglecting terms of order $m$
(which would be collinear finite) we can replace here $\vert
\underline{k}'\vert$ by $\vert \underline{p}'\vert \omega'/E_{p'}$.
Similarly the angle in $ \cos(\theta_{k'})$ may be replaced by
$\cos(\theta_{p'})$ since the correction is of the order of the
small angle $\delta$ and will only introduce finite corrections to
our integrals. In this way we see that, for the divergent terms, we
may rewrite  the numerator using
\begin{equation} p\scdot
k'=p\scdot p' \frac{\omega'}{E'}\,.
\end{equation}
This manipulation
allows us to express several divergent structures via a small number
of integrals.

\paragraph{Type 2 approximations}
We are interested in approximating $p_n\scdot k'$, where $p_n$ and
$k'$ are on-shell  but $p$ is not. We are interested in the region
where $p_n$ is almost collinear with $k'$ so these terms are small.
Naively, we might expect that the simple identity
\begin{equation}\label{angle20}
    p_n\scdot k'=(p-nk')\scdot k'=p\scdot k'\,,
\end{equation}
would suffice. However, in the text, we need to compare $p_n\scdot
k'$ with $p\scdot k'$ where $p$ is an on-shell momentum which is not
the case in~(\ref{angle20}) except in the exactly collinear (and
hence vanishing) limit. Thus a more careful argument is needed.

To proceed we need to relate the angles between the various vectors.
Let $\theta_n$ be the small angle between $p_n$ and $k'$, and let
$\theta$ be the small angle between $p$ and $\theta$. Recall that
$p_n=p-nk'$ and hence $p$ is not on-shell. Indeed, it is
straightforward to show that for small $\theta_n$ we have
\begin{equation}\label{angle21}
    p^2=nE_n\omega'\theta^2_n\,.
\end{equation}
Similarly, writing $p_n$ in terms of $p$ and $k'$, we see that the
on-shell condition for $p_n$ implies that
\begin{equation}\label{angle22}
    0=nE_n\omega'\theta^2_n-2nE\omega'+2n|\underline{p}|\omega'(1-\tfrac12\theta^2)\,.
\end{equation}
Now the spatial component of $p$  can be written as
$\underline{p}=\underline{p_n}+n\underline{k}$. Hence we can write
\begin{equation}\label{angle23}
|\underline{p}|=E-\frac{nE_n}{2E}\omega'\theta^2_n\,.
\end{equation}
Combining this result with (\ref{angle22}) we find the approximation
\begin{equation}\label{angle24}
    E_n^2\theta_n^2=E^2\theta^2\,.
\end{equation}
Using this result and the small angle approximation that
\begin{equation}\label{angle25}
    \frac1{p_n\scdot k'}=\frac{2}{\omega'
    E_n(\theta^2_n+\frac{m^2}{E_n^2})}\,,
\end{equation}
it is now straightforward to show that
\begin{equation}\label{angle26}
    \frac1{p_n\scdot k'}\frac{E}{E_n}=\frac1{p\scdot k'}\,.
\end{equation}

\section{An example of a dangerous argument}
In order to highlight the lack of convergence in summing the
degenerate processes that are essential for the Lee-Nauenberg
theorem, we will apply to the tree-level process the argument that
was used in~\cite{Akhoury:1997pb} at higher orders to try to prove
infrared finiteness for Coulomb scattering.

We start by considering the tree-level contribution $P_{0,0}$. This
is the basic process described by Figure~1.  At this order in
perturbation theory disconnected contributions can only arise due to
the introduction of disconnected soft photons. So, for example, in
$P_{1,1}$ the disconnected  photon lines contract upon themselves
resulting in the ill defined volume, denoted by~$\Box$, mentioned in
Section~7. So we can write $P_{1,1}=\Box P_{0,0}$. In a similar
manner, the combinatorics of contracting disconnected lines leads to
the identity $P_{2,2}=2(\Box^2+\Box)P_{0,0}$. In general, we see
that the Lee-Nauenberg probabilities that contribute to the \lq\lq
tree-level" cross-section are
\begin{equation}\label{asz4}
    P_{m,m}=\frac1{m!m!}D(m,m)P_{0,0}\,,
\end{equation}
where $D(m,m)$ are the disconnected contributions that arise from
$m$ straight through soft photon lines. So as we have seen
$D(0,0)=1$, $D(1,1):=\Box$ and  $D(2,2)=2(\Box^2+\Box)$. Note that,
just as in~\cite{Akhoury:1997pb}, we are not going to concern
ourselves with making sense of $\Box$.

The Lee-Nauenberg total cross-section is then
\begin{equation}\label{asz5}
    P=\sum_{m=0}^\infty P_{m,m}=\left(\sum_{m=0}^\infty\frac1{m!m!}D(m,m)\right)P_{0,0}\,.
\end{equation}
Now, if we apply the reasoning used in~\cite{Akhoury:1997pb}   to
this, we can derive something which is clearly incorrect.

First note the trivial identity
\begin{eqnarray}\label{asz6}
    \frac1{m!m!}D(m,m)&=&\sum_{a=0}^m\frac1{(m-a)!(m-a)!}D(m-a,m-a)\nonumber\\&&\qquad
    -\sum_{a=1}^m\frac1{(m-a)!(m-a)!}D(m-a,m-a)\,.
\end{eqnarray}
Here we are simply writing a term as a sum  minus the same sum
without the term. To make sense of this we must have $m>0$. Now we
write this identity as
\begin{eqnarray}\label{asz7}
    \frac1{m!m!}D(m,m)&=&\sum_{a=0}^m\frac1{(m-a)!(m-a)!}D(m-a,m-a)\\\nonumber&&\quad-
    \sum_{a=0}^{m-1}\frac1{(m-1-a)!(m-1-a)!}D(m-1-a,m-1-a)\,.
\end{eqnarray}
Hence, using the same  argument as in~\cite{Akhoury:1997pb}, we have
\begin{eqnarray}\label{asz8}
P&=&P_{0,0}+\sum_{m=1}^\infty P_{m,m}\nonumber\\
&=& P_{0,0} +\sum_{n=1}^\infty\frac1{m!m!}D(m,m)P_{0,0}\nonumber\\
&=&
P_{0,0}+\sum_{m=1}^\infty\sum_{a=0}^m\frac1{(m-a)!(m-a)!}D(m-a,m-a)P_{0,0}\\&&\quad-
\sum_{m=1}^\infty\sum_{a=0}^{m-1}\frac1{(m-1-a)!(m-1-a)!}D(m-1-a,m-1-a)P_{0,0}\,.\nonumber
\end{eqnarray}
Now we can absorb $P_{0,0}$ into the first sum and shift the $m$
counter in the second sum to $m-1$ to get
\begin{eqnarray}\label{asz9}
P &=&
\sum_{m=0}^\infty\sum_{a=0}^m\frac1{(m-a)!(m-a)!}D(m-a,m-a)P_{0,0}\nonumber\\&&\qquad-
\sum_{m=0}^\infty\sum_{a=0}^{m}\frac1{(m-a)!(m-a)!}D(m-a,m-a)P_{0,0}\,.
\end{eqnarray}
from which we can conclude that
\begin{equation}\label{asz10}
    P=\sum_{m=0}^\infty\sum_{a=0}^m\frac1{(m-a)!(m-a)!}D(m-a,m-a)(P_{0,0}-P_{0,0})=0\,.
\end{equation}
This is precisely the tree-level version of Equation~(14)
in~\cite{Akhoury:1997pb}. The implication, that there is no
tree-level scattering, shows that their argument is not safe!
\end{fmffile}
\newpage


\end{document}